\documentclass[a4paper,11pt]{article}
\pdfoutput=1 

\usepackage{jcappub} 

\usepackage[T1]{fontenc} 
\usepackage{amsmath}
\usepackage{amsfonts}
\usepackage{amsthm}
\usepackage{amssymb}
\usepackage{latexsym, array,multirow,  verbatim, enumerate}
\usepackage{slashed}
\usepackage{booktabs}
\usepackage{tabularx}
\usepackage{cancel}
\usepackage{dcolumn}
\usepackage{bm}
\usepackage{graphicx, caption, subcaption} 
\usepackage{url}
\usepackage[compat=1.0.0]{tikz-feynman}
\usepackage{pifont}
\usepackage{empheq}
\usepackage[utf8]{inputenc}
\usepackage{bm}
\usepackage{blkarray}
\usepackage[font=footnotesize,labelfont=sf]{caption}
\usepackage{cleveref}
\usepackage{float}
\usepackage{multirow}
\newcommand{\hquad}{\hspace{3em}}
\newcommand{\solarmass}{\textup{M}_\odot}

\title{\boldmath Exploring millicharged dark matter components from the shadows }

\author[a,1]{Lalit S. Bhandari\note{Corresponding author.}}
\author[a]{and Arun M. Thalapillil}

\affiliation[a]{ Department of Physics,\\ Indian Institute of Science Education and Research Pune,\\ Pune 411008, India}

\emailAdd{bhandari.lalitsingh@students.iiserpune.ac.in}
\emailAdd{thalapillil@iiserpune.ac.in}

\abstract{ Dark matter sectors with hidden interactions have been of much interest in recent years. These frameworks include models of millicharged particles as well as dark sector bound states, whose constituents have electromagnetic gauge interactions. These exotic, charged states could constitute a part of the total dark matter density. In this work, we explore in some detail the various effects, on the photon sphere and shadow of spherically symmetric black holes, due to dark matter plasmas furnished by such sectors. Estimating physically viable parameter spaces for the particle physics models and taking semi-realistic astrophysical scenarios that are amenable to theoretical analyses, we point out various modifications and characteristics that may be present. Many of these effects are unique and very distinct from analogous situations with conventional baryonic plasmas, or neutral perfect fluid dark matter surrounding black holes. While in many physically viable regions of the parameter space the effects on the near-horizon regions and black hole shadows are small, in many parts of the low particle mass regions the effects are significant, and potentially measurable by current and future telescopes. Such deviations, for instance, include characteristic changes in the photon sphere and black hole shadow radii, unique thresholds for the dark matter plasma dispersion where the photon sphere or black hole shadow vanishes, and  where the dark matter plasma becomes opaque to electromagnetic waves. Alternatively, we point out that a non-observation of such deviations and characteristics, in future, could put constraints on interesting regions of the particle physics parameter space.}

\begin{document}
\maketitle
\flushbottom

%%%%%%%%%%%%%%%%%%%%%%%%%%%%%%%%%%%%%%%

\section{Introduction}
\label{sec:intro}
The presence of dark matter (DM) in the universe is among the foremost outstanding problems in physics today. Among the viable candidates that have been proposed as constituting this DM density, in part or in full, are millicharged particles (mCP)\,\cite{Holdom:1985ag,Goldberg:1986nk, Cheung:2007ut,Kors:2004dx,Feldman:2007wj,Dienes:1996zr,Abel:2003ue,Batell:2005wa} and bound states arising from dark sectors that mirror the Standard Model\,\cite{Kobzarev:1966qya, Okun:2006eb,Foot:2014mia,Kaplan_2010,Kaplan_2011}. 

There have been extensive studies on DM with self-interactions (see for instance\,\cite{Tulin:2017ara}, and references therein), motivated by various astrophysical and cosmological conundrums\,\cite{Spergel:1999mh}. In many cases, these self-interacting components have also been shown to lead to unique DM structures in the universe\,\cite{Fan:2013tia,Fan:2013yva}. DM with a hidden dark charge and constraints on them from cosmological and astrophysical observations have also been studied\,\cite{Feng:2009mn}. As we mentioned above, in this broad category, mCP is a theoretically well-motivated candidate\,\cite{Holdom:1985ag,Goldberg:1986nk, Cheung:2007ut,Kors:2004dx,Feldman:2007wj,Dienes:1996zr,Abel:2003ue,Batell:2005wa} with interesting signatures and constraints (see for example,\,\cite{Jaeckel:2010ni,Collar:2012olx}). Exotic particles carrying millicharges are also of much current interest---as possible explanations for various anomalies, and as prime candidates for focused searches (see for instance some of the recent, representative works in\,\cite{Li:2021zcy,Aboubrahim:2021ohe,Berlin:2021kcm,Budker:2021quh,Bai:2021nai,ArguellesDelgado:2021lek,Munoz:2021lxp,Foroughi-Abari:2020qar,Plestid:2020kdm}). 

On the other hand, based on overwhelming and varied observational evidence\,\cite{Abbott:2016blz,1997MNRAS.284..576E,1998ApJ...509..678G,Akiyama:2019cqa}, it is now generally believed that black holes are ubiquitous objects in the universe and that almost all galaxies have supermassive black holes at their centre. Along with the observation of gravitational waves\,\cite{Abbott:2016blz}, the observation of the near-horizon region of a supermassive black hole\,\cite{Akiyama:2019cqa} has also ushered in a new era for probing gravitational, astrophysical and fundamental physics questions using these systems. 

Black hole shadows\,\cite{1974IAUS...64..132B} have been extensively studied for various spacetime backgrounds (See for instance\,\cite{Cunha:2018acu, Perlick:2021aok}, and references therein). The effects on the black hole characteristics due to a modification of the spacetime background by any surrounding dark matter or exotic fields has also received much attention in recent years (see\,\cite{Arvanitaki:2009fg, Brito:2015oca, Cunha:2015yba,Hou:2018bar,Konoplya:2019sns,Jusufi:2019nrn} and related references, for instance). In particular, there have been a few recent studies\,\cite{Xu:2017bpz,Haroon:2018ryd,Hou:2018avu,Jusufi:2019ltj,Shaymatov:2020bso,Saurabh:2020zqg,Badia:2020pnh,Rayimbaev:2021kjs,Atamurotov:2021hck} that have considered effects due to a perfect fluid DM, with a special equation of state\,\cite{Kiselev:2003ah,Li:2012zx} that is quintessential or phantom field like with $P_r\simeq-\rho$, surrounding a black hole. In contrast to these studies, our focus in this work will be on studying the effects due to a DM plasma, furnished by mCP components, taking phenomenologically viable mCP and semi-realistic astrophysical scenarios. 

There have also been seminal studies on the influence of hydrogen plasma surrounding black holes\,\cite{perlick2000ray,PhysRevD.92.104031,Chowdhuri:2020ipb,li2021gravitational,Badia:2021kpk}. For realistic parameters, the effect of a hydrogenic plasma environment on the spherically symmetric black hole shadow radius is found to be unimportant\,\cite{PhysRevD.92.104031,li2021gravitational}. Recently, there has also been a study on baryonic plasma effects on gravitational lensing by static black holes, in the presence of a phantom field like perfect fluid\,\cite{Atamurotov:2021hoq}. Understanding the conventional physics around such extreme objects, as well as effects of any new physics in their neighbourhoods, engendered by the extreme environments, may open new avenues for discovery. As we will see, for a plasma component sourced by mCP particles, the change in both the photon sphere and shadow radii may be very significant, and much higher than that of baryonic plasmas.

In this work, our interest is in exploring what effects mCP sectors may have on the near horizon regions of spherically symmetric black holes and their corresponding shadows. If mCP species form a component of DM, there is potential for this DM plasma-like component to leave an imprint on the photon sphere and shadow radii. The enormous gravitational well created by black holes may serve to locally amplify DM densities, further amplifying potential effects. One such well-studied astrophysical scenario, for instance, is the formation of DM spikes\,\cite{Gondolo_1999,Lacroix2015} near black holes. The consequences of such local DM over-densities have received some attention lately\,\cite{Shelton:2015aqa,nampalliwar2021modelling,Feng:2021qkj,Kim:2021yyo}. We wish to study in detail the theoretical and phenomenological effects of a mCP sourced DM plasma in such frameworks. Towards this aim, we derive the relevant theoretical expressions and point out the salient features in various simple, but semi-realistic, astrophysical scenarios and physically viable regions of the mCP parameter space. We adapt and generalise the methods developed for baryonic plasmas\,\cite{perlick2000ray,PhysRevD.92.104031,Chowdhuri:2020ipb,li2021gravitational,Badia:2021kpk} in general relativity, to study the effect of this charged DM fluid component, furnished by the mCPs (henceforth referred to as DM plasma). 

The main results we obtain are that while in most regions of the physically viable mCP parameter space the effects on the photon sphere and black hole shadow radii are small, there are extensive regions where the effects are significant and potentially observable. We first re-derive the possible DM ionisation fraction, for the case of interest to us, to determine the extent of DM plasma allowed, as a constituent in the total DM content. We adapt methodologies from previous seminal works\,\cite{2004,Feng:2009mn,PhysRevD.85.101302}, to perform these estimates for the mCP parameter space of relevance to the study. We then consider two simple cases for the DM plasma distribution, that of constant density and that of radially in-falling DM plasma. 

In the former case we find that the variation in the shadow radius may be as high as $\sim25\%$. There are also concomitant characteristic variations in the photon sphere and shadow radii, as the effect of the DM plasma increases; within the physically viable mCP parameter space. Among the other interesting features are the presence of threshold values where the photon sphere disappears and the plasma becomes opaque to the electromagnetic radiation. In the latter scenario, with in-falling DM plasma, there are again distinct changes in the photon sphere and black hole shadow in interesting regions of the mCP parameter space. There is a threshold value now where the photon sphere grazing light trajectory is attenuated and also remarkably where, due to the refractive effect of the DM plasma, the black hole shadow technically disappears from the observer's perspective. Apart from the theoretical expressions and features that we quantify and describe, some of the main results are those summarised in Figs.\,\ref{fig:CMDchid}-\ref{fig:cdm_rif_constraint}, and Tables\,\ref{tab:mCPAPCDem}-\ref{tab:mCPAsheffir}.

In Sec.\,\ref{sec:mcps} we briefly review the theoretical framework and current constraints on mCP states. Here, we also adapt methods from earlier studies to place relevant constraints and estimate limits on the DM plasma component, for the parameter space of interest to us. In Sec.\,\ref{sec:bhs} we introduce the relevant formalism for calculating black shadows. Sec.\,\ref{sec:mcpshdw} contains the main analyses and results, where we study the DM plasma effects in detail. We summarise the key findings of the study and conclude in Sec.\,\ref{sec:summary}.
%%%%%%%%%%%%%%%%%%%%%%%%%%%%%%%%%%%%%%%

%%%%%%%%%%%%%%%%%%%%%%%%%%%%%%%%%%%%%%%

\section{Millicharged particles}
\label{sec:mcps}
As we motivated in the introduction, components of DM with self interactions are interesting  from diverse viewpoints. mCPs are interesting candidates in this category and have been studied extensively\,\cite{Holdom:1985ag,Goldberg:1986nk, Cheung:2007ut,Kors:2004dx,Feldman:2007wj,Dienes:1996zr,Abel:2003ue,Batell:2005wa}. They may even potentially constitute a component of the DM in the universe in some form\,\cite{Goldberg:1986nk,Feng:2009mn,Kaplan_2010,Kaplan_2011,PhysRevD.85.101302,Cline:2021itd}. In this section, to set the context and clarify notations, we briefly review the theoretical framework and constraints on mCPs. We also reformulate some of the derivations, limits and discussions for the region of parameter space that will be of interest to us in Sec.\,\ref{sec:mcpshdw}. We set $\hbar=1$ everywhere.
%%%%%%%%%%%%%%%%%%%
\subsection{Theoretical framework and constraints for millicharged particles}
\label{subsec:tfcmcps}
mCPs arise naturally in many theoretical models. One of the simplest frameworks is one that contains at least two $U(1)$ gauge groups---one corresponding to the visible sector, and the other corresponding to a dark sector, with Standard Model gauge singlet particles. The interaction between the visible and dark sector may be mediated, for instance, by a kinetic mixing\,\cite{Holdom:1985ag} portal. This leads to Lagrangian terms, in the simplest case, of the form
\begin{equation}\label{eq:flTFMCP}
\mathcal{L}\supset-\frac{1}{4}F^{\mu\nu}F_{\mu\nu}-\frac{1}{4}\tilde{F}^{\mu\nu}\tilde{F}_{\mu\nu}-\frac{\chi}{2}F^{\mu\nu}\tilde{F}_{\mu\nu} \;.
\end{equation}
Here, $F^{\mu\nu}=\partial^{\mu}A^{\nu}-\partial^{\nu}A^{\mu}$ and $\tilde{F}^{\mu\nu}=\partial^{\mu}\tilde{A}^{\nu}-\partial^{\nu}\tilde{A}^{\mu}$  are the field tensors, with  $A^\mu$ and  $\tilde{A}^\mu$ denoting the Standard Model and dark sector $U(1)$ gauge fields respectively. Both gauge fields are assumed to be massless. The last term in Eq.\,(\ref{eq:flTFMCP}) is the kinetic mixing term\,\cite{Holdom:1985ag}. Such a term may be generated at high energy scales, via loop interactions involving heavy messenger particles\,\cite{Holdom:1985ag} or through non-perturbative physics\,\cite{DelZotto:2016fju}. Another alternative for the coupling is through St\"{u}ckelberg mixing\,\cite{Kors:2004dx, Feldman:2007wj}, with an additional St\"{u}ckelberg field that transforms in a particular way under gauge transformations. Millicharged particles may also arise directly, in extra-dimension frameworks, by separating fermions in the warped bulk from a localized visible sector $U(1)$ gauge boson\,\cite{Batell:2005wa}. Therefore, $\chi$ should be generically thought of as a real-valued parameter.

The gauge bosons will in general have couplings to matter fields. Consider, for instance, a single species of dark sector fermion ($\tilde{\psi}$), with a charge $\tilde{q}$ and mass $m_{\tilde{\psi}}$ . It will have a coupling to the dark gauge field through the kinetic term
\begin{equation}\label{eq:cilTFMCP}
\mathcal{L}^{\text{\tiny f}}\supset \, \bar{\tilde{\psi}}\left(i\slashed{D}-m_{\tilde{\psi}}\right)\tilde{\psi}\;,
\end{equation}
where ${D_{\mu}}={\partial_{\mu}}+i\tilde{q}\tilde{{A}}_{\mu}$ is the covariant derivative and $\slashed{D}=\gamma^\mu D_\mu$. To obtain the physical gauge fields, we must require canonical normalization of the gauge field kinetic terms, i.e. the kinetic term has to be diagonalised. For $\chi\ll 1$, up to leading order in $\chi$,  this may be achieved by the field redefinition~\cite{Holdom:1985ag} $\tilde{A}_{\mu} \to \tilde{A}_{\mu}- \chi {A}_{\mu}$. Under this, Eq.\,(\ref{eq:cilTFMCP}) now becomes
\begin{equation}\label{eq:ncilTFMCP}
\mathcal{L}^{\text{\tiny f}} \supset \bar{\tilde{\psi}}\left(i\slashed{D}+\chi\tilde{q}{\slashed{A}}-m_{\tilde{\psi}}\right)\tilde{\psi}\;.
\end{equation}
We see that under this transformation, though the redefined physical dark gauge field still couples only with the dark fermions, the visible sector physical gauge field now has couplings to both the Standard Model fermions and the dark sector fermion.

If, alternatively, the dark sector matter fields were scalar particles ($\tilde{\phi}$) with a charge $\tilde{q}$ and mass $m_{\tilde{\phi}}$, the interaction terms would arise from the kinetic term
\begin{equation}\label{eq:cilTSMCP}
\mathcal{L}^{\text{\tiny s}} \supset \left(D_\mu\tilde{\phi}\right)^\dagger \left(D^\mu\tilde{\phi}\right)-m^2_{\tilde{\phi}}|\tilde{\phi}|^2 \; .
\end{equation}
Again, for $\chi\ll 1$, under the  field redefinition $\tilde{A}_{\mu} \to \tilde{A}_{\mu}- \chi {A}_{\mu}$, making the gauge field kinetic terms canonical, Eq.\,(\ref{eq:cilTSMCP}) becomes
\begin{equation}\label{eq:ncilTSMCP}
\mathcal{L}^{\text{\tiny s}} \supset\left(D_\mu\tilde{\phi}-i\chi\tilde{q}{{A}}^\mu\tilde{\phi}\right)^\dagger \left(D^\mu\tilde{\phi}-i\chi\tilde{q}{{A}}^\mu\tilde{\phi}\right)-m^2_{\tilde{\phi}}|\tilde{\phi}|^2\;.
\end{equation}
As before, the visible sector physical gauge boson develops a coupling with the dark sector scalar particle. 

 We will refer to the physical gauge fields, after canonical normalisation of the gauge kinetic terms, as the photon and the dark photon henceforth. We see from above that the dark sector particle will be effectively seen as a fractionally charged particle, by the visible sector photon, with electromagnetic charge 
\begin{equation}\label{eq:epsilon}
\epsilon=\frac{\chi \tilde{q}}{q_e}\; .
\end{equation}
The dark sector particle now forms the mCP. Note that the charge ($\epsilon$) of the mCP is being expressed in units of the electron charge magnitude ($q_e$). As mentioned earlier, mCPs may also arise in extra-dimension frameworks as manifestations of fermions separated in the warped bulk from a localized visible sector $U(1)$ gauge boson\,\cite{Batell:2005wa}, in which case there is no dark gauge interaction necessarily between the mCPs.
%%%%%%%%%%%%
\begin{figure}[h!]
	\centering{\includegraphics[scale=0.235]{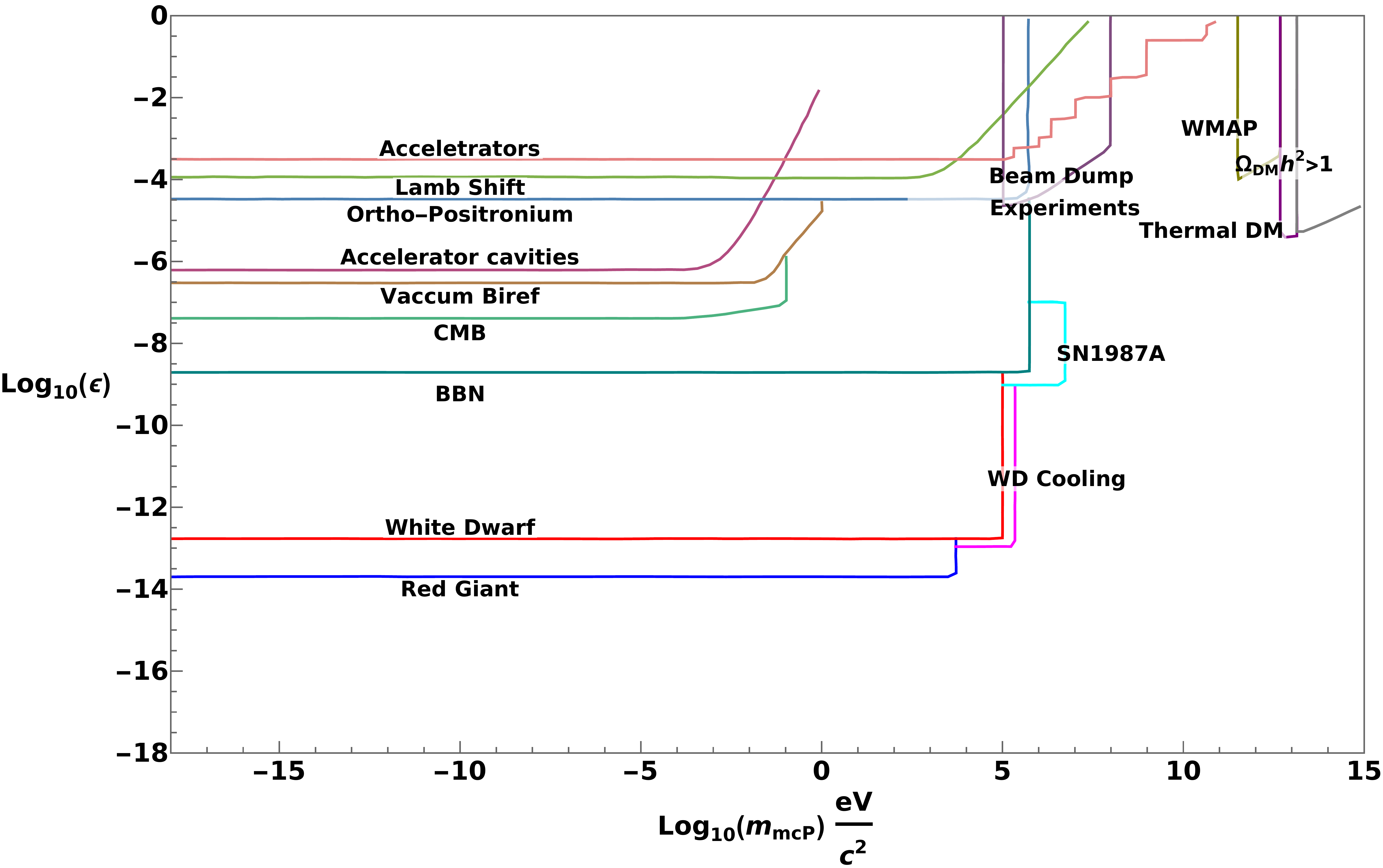}}				
	\caption{ Current constraints on mCPs from various astrophysical, cosmological and terrestrial experiments and considerations\,\cite{essig2013dark,Vogel_2014,Mohapatra:1990vq,Dubovsky_2004,PhysRevD.43.2314,Davidson_2000,PhysRevLett.81.1175,PhysRevD.35.391}. For the low-mass regions, the most stringent bounds come from Red Giant and White Dwarf cooling limits, restricting the sub-eV mCP charge to $\epsilon\lesssim 10^{-14}$.}
	\label{fig:mcPconstraint}
\end{figure}
%%%%%%%%%%%

For completeness, we note that to arbitrary order in $\chi$, diagonalization can also be acheived~\cite{2021}. Such a treatment shows that mCP charge comes out to be $\epsilon q_e=\chi \tilde{q}/\sqrt{1-\chi^2} $. For small $\chi$, this agrees with the simpler derivation leading to Eq.\,(\ref{eq:epsilon}). We also mention that with a massless, dark sector gauge field in Eq.\,(\ref{eq:flTFMCP}) we have the freedom to choose arbitrary linear combinations of the gauge fields as the mass eigenstates. The relevant physics though remains the same.

Let us now state very briefly some of the main constraints that are present for mCPs. Many regions of the $\epsilon-m_{\text{\tiny{mCP}}}$ parameter space are constrained by various terrestrial, astrophysical and cosmological considerations (See\,\cite{essig2013dark,Vogel_2014,Mohapatra:1990vq,Dubovsky_2004,PhysRevD.43.2314,Davidson_2000,PhysRevLett.81.1175,PhysRevD.35.391} and references therein). The various constraints based on these observations and studies are summarised in Fig.\,\ref{fig:mcPconstraint}.

For the low-mass region ($m_{\text{\tiny{mCP}}} \ll \mathcal{O}(1)\,\mathrm{eV}$) of the mCP parameter space, for instance, the most stringent constraints are from astrophysics and cosmology. Assuming that the mCPs were in thermal equilibrium with the rest of matter during the early stage of universe, they can effect the abundance of elements. This contribution of the mCPs to the Big-Bang nucleosynthesis (see for instance\,\cite{PhysRevD.43.2314} and related references) phenomenology and comparison with the observed abundances suggest the allowed regions are around
\begin{equation}
\epsilon\lesssim 2 \times 10^{-9}\; ,
\end{equation}
for $m_{\text{\tiny mcp}}\lesssim 200\,\mathrm{ MeV}$. The most exacting astrophysical bounds in this region though come from considerations of Red Giant and White Dwarf cooling (Again, see for instance\,\cite{PhysRevD.43.2314} and related references). The presence of mCPs, due to plasmon decay in the stellar plasma, would increase the cooling rate of Red Giants and White Dwarfs by removing energy from the stellar cores. This would alter their normal evolutions. These studies\,\cite{PhysRevD.43.2314,Davidson_2000,Vogel_2014}  give the viable region as 
\begin{equation}\label{eq:rgwdlim}
\epsilon \lesssim 10^{-13}-10^{-14}\;,
\end{equation} 
when  $m_{\text{\tiny{mCP}}} < \mathcal{O}(1)\,\mathrm{eV}$.

As has been pointed out though, the bounds from astrophysics may be relaxed and evaded in many cases\,\cite{Masso:2006gc, Abel:2006qt, Foot:2007cq, Melchiorri:2007sq,DeRocco:2020xdt}. For instance, if one allows for multiple dark $U(1)_{\text{\tiny{D}}}$ gauge groups, and associated dark photons, one may have for the effective mCP charge in the astrophysical plasma\,\cite{Masso:2006gc}
\begin{equation}
Q_{\text{\tiny{mCP}}}(q^2\rightarrow \omega_{\text{\tiny P}}^2) \ll Q_{\text{\tiny{mCP}}}(q^2\rightarrow 0)\equiv \epsilon \; .
\end{equation}
Here, $q$ is the typical momentum transfer between the charged particles in the medium and $\omega_{\text{\tiny P}}^2$ is the plasma frequency in the corresponding astrophysical environment. Thus, the effective mCP charge in the stellar plasma could be very small, satisfying the cooling bounds in Eq.\,(\ref{eq:rgwdlim}), while in near-vacuum having mCP charges  $\epsilon\sim 10^{-7}$ or larger\,\cite{Masso:2006gc, Abel:2006qt, Foot:2007cq, Melchiorri:2007sq}. For light fermionic mCPs, this loop-hole may be mitigated and strong limits placed, in some parameter space regions, by considering Schwinger pair production in Magnetar environments\,\cite{Korwar:2017dio}. 

%%%%%%%%%%%%%%%%%%%
\subsection{Parameter space of interest for black hole shadow analyses}
\label{subsec:imps}
In our study, the main effects on the black hole shadow will be due to the dispersive effect induced by charged mCPs. We are considering the mCPs as subcomponents of the total DM. Assume then that $f_\text{\tiny mCP}$ denotes the mass fraction of DM comprised of mCPs. We would like to explore what fraction of the total DM density may be comprised in mCPs, in regions of the $\epsilon-m_{\text{\tiny{mCP}}}$ parameter space where the effects on the black hole shadow may in principle be significant.

We will see in Sec.\,\ref{sec:mcpshdw} that the most significant changes in the shadow radius come from the sub-eV mass scale for mCPs. In these sub-eV regions, if two different mCP species (say $\tilde{e}$ and $\tilde{p}$) exist with opposite charges, bound states may not always necessarily form, as the Bohr radii are sometimes much higher than the inter-particle distances. Also, the corresponding binding energies are much smaller than the ambient energies. Therefore, in these regions, the $\tilde{e}$ and $\tilde{p}$ would exist mostly in their unbound states. Note that the naive Tremaine-Gunn bound\,\cite{PhysRevLett.42.407} would constrain light fermionic DM components to have a mass $m^f_\text{\tiny mCP} \gtrsim\mathcal{O}(100)\,\mathrm{eV}$. However it has been pointed out that such constraints may be easily evaded by having a large number of species\,\cite{Davoudiasl:2020uig}, and masses $m^f_\text{\tiny mCP} \gtrsim\mathcal{O}(10^{-14})\,\mathrm{eV}$ may be possible. Nevertheless, to be concrete, we will conservatively assume our mCP species to be scalar particles. We will however continue to refer to these scalar mCP species as  $\tilde{e}$ and $\tilde{p}$.

An important question, in this context, is then to ascertain what the constraints on the fraction of these ionised charged components are in the total DM distribution. One of the major constraints on DM fractions that are in an ionised state are from investigations of the Bullet cluster\,\cite{Feng:2009mn}. They are based on considerations of the mass fraction lost by the subcluster, when it passed through the main cluster. Additional interactions between DM particles would effect the mass lost and the observations can thereby put limits on them. Specifically, due to the additional interaction, the particles from the main cluster and the subcluster may acquire a velocity greater than the escape velocity of the subcluster after collision. For this scenario to occur, the scattering angle in the subcluster's reference frame should lie in the range~\cite{2004}
 \begin{equation}
 	\frac{v_{\text{\tiny esc}}}{v}< \cos\theta<\sqrt{1-\left(\frac{v_{\text{\tiny esc}}}{v}\right)^2}\;.
 \end{equation}
 Here, $v$ is the final velocity of the particle, $v_{\text{\tiny esc}}$ is the escape velocity of the subcluster and $\theta$ is scattering angle, all in the subcluster's frame of reference. 
 
  In\,\cite{Feng:2009mn}, limits on DM self-interactions were studied, with a dark coupling constant $\tilde{\alpha}$, assuming all of DM has self interactions. We adapt the methodology in\,\cite{Feng:2009mn} for the sub-eV mCP mass range of interest, including both an $\bar{\alpha}=\text{max}(\tilde{\alpha},\epsilon^2\alpha)$ and allowing for the possibility of a mass fraction $ f_\text{\tiny mCP}$, that is in an ionised state. Ion-ion interactions may be modelled by Rutherford scattering. The differential cross section for mCP-mCP scattering may be written in our case as
 \begin{equation}
 	\frac{d \sigma}{d \Omega}=\frac{\bar{\alpha}^2\hbar^2 c^2}{4 m^2_{\text{\tiny mCP}}v^4\sin^4(\theta/2)}\;,
 \end{equation}
 where $\bar{\alpha}=\text{max}(\tilde{\alpha},\epsilon^2\alpha)$, as before. Here, $\alpha$ and $\tilde{\alpha}$ are the visible and dark sector fine structure constants.
 
If $\Sigma_{s}$ is the dark matter surface density of the subcluster, then the surface number density of mCPs in the subcluster will be
 \begin{equation}
 	n_{\text{\tiny mCP}}=f_\text{\tiny mCP}\frac{\Sigma_{s}}{m_{\text{\tiny mCP}}} \,.
 \end{equation}
 
 The probability that a collision of a particle from the main cluster and that from the subcluster will result in both getting knocked out is given by the expression
 \begin{equation}
 	\mathcal{P}\equiv f_\text{\tiny mCP}\frac{\Sigma_{s}}{m_{\text{\tiny mCP}}}\int_{\theta_{\text{\tiny min}}}^{\theta_{\text{\tiny max}}}d\Omega\frac{d \sigma}{d \Omega}=\frac{2\pi f_\text{\tiny mCP}\Sigma_{s}\bar{\alpha}^2\hbar^2 c^2}{ m^3_{\text{\tiny mCP}}v^4}\left(\frac{1}{1-\cos\theta_{\text{\tiny min}}}-\frac{1}{1-\cos\theta_{\text{\tiny max}}}\right)\,.
 \end{equation}
 We are neglecting any screening effects. Following the study of \cite{2004} (Galaxy cluster 1E 0657-56, Bullet cluster), we take the values for the  total dark matter surface density of the subcluster as $\Sigma_{s}\simeq 0.3\,\mathrm{g/cm^2}$, velocity of the particle in the main cluster as $v\simeq 4800\,\mathrm{Km/s}$ and the escape velocity as $v_{\text{\tiny esc}}\simeq 1200\,\mathrm{Km/s}$. Furthermore, we assume that not more than $25\%-50\%$ of the mCPs in the subcluster are lost in the collision (i.e. $\mathcal{P}\le 0.25-0.5$). Using the parameters above, we then then obtain a limit 
 \begin{equation} \label{eq:bcc}
 f_\text{\tiny mCP}\,\bar{\alpha}^2\lesssim 10^{-33}\left(\frac{m_\text{\tiny mCP}}{\text{ eV}}\right)^3 \;.
 \end{equation} 

Eq.\,(\ref{eq:bcc}) gives an upper limit for $f_\text{\tiny mCP}$, for a given value of $\bar{\alpha}$ and $m_\text{\tiny mCP}$. For instance, assuming  $\tilde{\alpha}\sim \epsilon^2\alpha$, if $\epsilon\sim10^{-14}$ and $m_{\text{\tiny mCP}}\sim 10^{-10}\,\mathrm{eV}$, it is found that $f_{\text{\tiny mCP}}\le(0.2-0.5)\%$. If $\epsilon\sim10^{-18}$ and $m_{\text{\tiny mCP}}\sim 10^{-15}\,\mathrm{eV}$, one would alternatively obtain $f_{\text{\tiny mCP}}\le(2-5)\%$. For a cold dark sector, assuming the bounds in Eqs.\,(\ref{eq:rgwdlim}) and (\ref{eq:bcc}) will also satisfy the CMB and BBN constraints\,\cite{Vogel_2014,Cyr-Racine:2013fsa, Fan:2013tia,Fan:2013yva}. In the study we will conservatively take $f_{\text{\tiny mCP}}=0.1\%$, as any neglected screening effects due to the oppositely charged $\tilde{e}$ and $\tilde{p}$ species will only increase the allowed fraction.

Though for most regions of the mCP parameter space this relative mass fraction is small, local DM mass density fluctuations around black holes may still lead to large local mCP mass densities in some cases. One such mechanism is the formation of DM spikes\,\cite{Gondolo_1999,Lacroix2015}, near black holes, as we already mentioned in the introduction. DM spikes may lead to peak densities around a black hole as high as $\sim 10^{11}\,\mathrm{GeV}\mathrm{/cm}^3$ and average DM densities exceeding normal expectations\,\cite{Gondolo_1999,Lacroix2015}. We will leverage this prospect in Sec.\,\ref{sec:mcpshdw} and consider the possibility of local enhancements in DM densities. However, we will only assume DM plasma densities that are much smaller than those allowed by DM spikes, and thereby be still conservative in our estimates.

%%%%%%%%%%%%%%%%%%%%%%%%%%%%%%%%%%%%%%%%%%%%%%%%%%%%%%%%%%%%%
\section{Black hole shadows}
\label{sec:bhs}
Consider the static, spherically symmetric exterior geometry described by the metric
\begin{equation}\label{eq:SSGF}
ds^2=-f(r)dt^2+\frac{dr^2}{g(r)}+r^2d\Omega^2\; .
\end{equation}
$(r,\theta,\phi)$ are spherical coordinates and $d\Omega^2=d\theta^2+\sin^2\theta d\phi^2$ is the solid angle element. 

The static and spherically symmetric nature of the spacetime furnishes corresponding Killing vectors ($K^\mu$) related to these isometries. They satisfy $\nabla_{\{\alpha} K_{\beta\}}=0$ ($\{\}$ denoting symmetrisation with respect to the appropriate indices) and imply that the quantity $K^\alpha p_\alpha$ associated with them is conserved along the geodesic. We will be primarily interested in null geodesics, and a suitable normalization of the affine parameter $\zeta$ lets us write the conserved quantities as $g_{\alpha\beta} K^\alpha dx^\beta/d\zeta$. The time-like killing vector $\partial_t$ gives the conservation of energy and the space-like vector $\partial_\phi$ leads to the conservation of angular momentum
\begin{eqnarray}\label{eq:CQ}
E=f(r)\frac{dt}{d\zeta}\;,\nonumber \\
L=r^2\frac{d\phi}{d\zeta} \; .
\end{eqnarray}
Metric compatibility and the geodesic equation for the photon trajectory (lightlike) also imply that
\begin{equation}\label{eq:leconst}
\frac{1}{2}\left(\frac{dr}{d\zeta}\right)^2+\frac{g(r)}{2}\left[\frac{L^2}{r^2}-\frac{E^2}{f(r)}\right]=0\; ,
\end{equation}
along the null geodesic. This may be viewed as an equation of the form
\begin{equation}\label{eq:RDE}
\frac{1}{2}\left(\frac{dr}{d\zeta}\right)^2+\tilde{V}_{eff}=0\;,
\end{equation}  
with an effective potential
\begin{equation}\label{eq:Ve}
\tilde{V}_{eff}=\frac{g(r)}{2}\left[\frac{L^2}{r^2}-\frac{E^2}{f(r)}\right]\;.
\end{equation}
Eq.\,(\ref{eq:leconst}) may also be re-arranged, using Eq.\,(\ref{eq:CQ}), to give
\begin{equation}\label{eq:RPHI}
\frac{dr}{d\phi}=\pm r\sqrt{g(r)\left[\frac{E^2r^2}{L^2f(r)}-1\right]} \; .
\end{equation}
The above expression gives the photon trajectories and orbit equations in the spacetime background. The light ray trajectories depend on both the $g_{tt}$ and $g_{rr}$ metric components (also see Appendix A). Note that the above equation imposes a condition
\begin{equation}
\frac{E^2}{L^2}\ge\frac{f(r)}{r^2} \quad \forall \; r \;.
\end{equation}
Of particular importance to us, in the context of the study, will be unstable circular orbits around the black hole and trajectories that graze these unstable orbits.

The critical points ($r_c$) of interest to us are the turning points of the orbit. At these points
 \begin{equation}\label{eq:tpc}
\left.\frac{dr}{d\phi}\right|_{r=r_c}=0 \equiv \tilde{V}_{eff}\left(r=r_c\right)=0 \; ,
\end{equation}
and from Eq.\,(\ref{eq:RPHI}), this gives
\begin{equation}
\frac{E}{L}=\frac{\sqrt{f(r_c)}}{r_c}\;.
\end{equation}
Using this, we may rewrite Eq.\,(\ref{eq:RPHI}) in terms of the critical points as
\begin{equation}\label{eq:RPHI2}
\frac{dr}{d\phi}=\pm r\sqrt{g(r)\left[\frac{f(r_c)r^2}{f(r)r_c^2}-1\right]}\;.
\end{equation}

The photon sphere denotes the smallest inner circular orbit for the photons. The required criteria for an unstable photon sphere, from Eq.\,(\ref{eq:RDE}), are
\begin{equation}\label{eq:Vefc}
\tilde{V}_{eff}(r_{\text{\tiny{ph.}}})=0~~,~~~\tilde{V}'_{eff}(r_{\text{\tiny{ph.}}})=0 ~~,~~~~ \tilde{V}''_{eff}(r_{\text{\tiny{ph.}}})<0 \;.
\end{equation} 
These conditions immediately lead to an implicit equation for the photon sphere radius, of the form
\begin{equation}\label{eq:COC}
r_{\text{\tiny{ph.}}}f'(r_{\text{\tiny{ph.}}})=2f(r_{\text{\tiny{ph.}}}) \; .
\end{equation}

For the case of the static, spherically symmetric, vacuum black hole exterior solution---the Schwarzschild metric---we have
\begin{equation}\label{eq:SF}
f(r)^{\text{\tiny{Sch.}}}=g(r)^{\text{\tiny{Sch.}}}=1-\frac{2M}{r}\;,
\end{equation} 
and we get from Eq.\,(\ref{eq:COC}) the familiar result
\begin{equation}\label{eq:SGUO}
r^{\text{\tiny{Sch.}}}_{\text{\tiny{ph.}}}=3M\;.
\end{equation}
In this case, as is well-known, Eq.\,(\ref{eq:SGUO}) is the only solution and it is unstable with
 \begin{equation}\label{eq:SGV''}
\left.\tilde{V}''_{eff}\right|_{r=r^{\text{\tiny{Sch.}}}_{\text{\tiny{ph.}}}}<0\;.
\end{equation}

\begin{figure}[h]
\center
\includegraphics[scale=1.05]{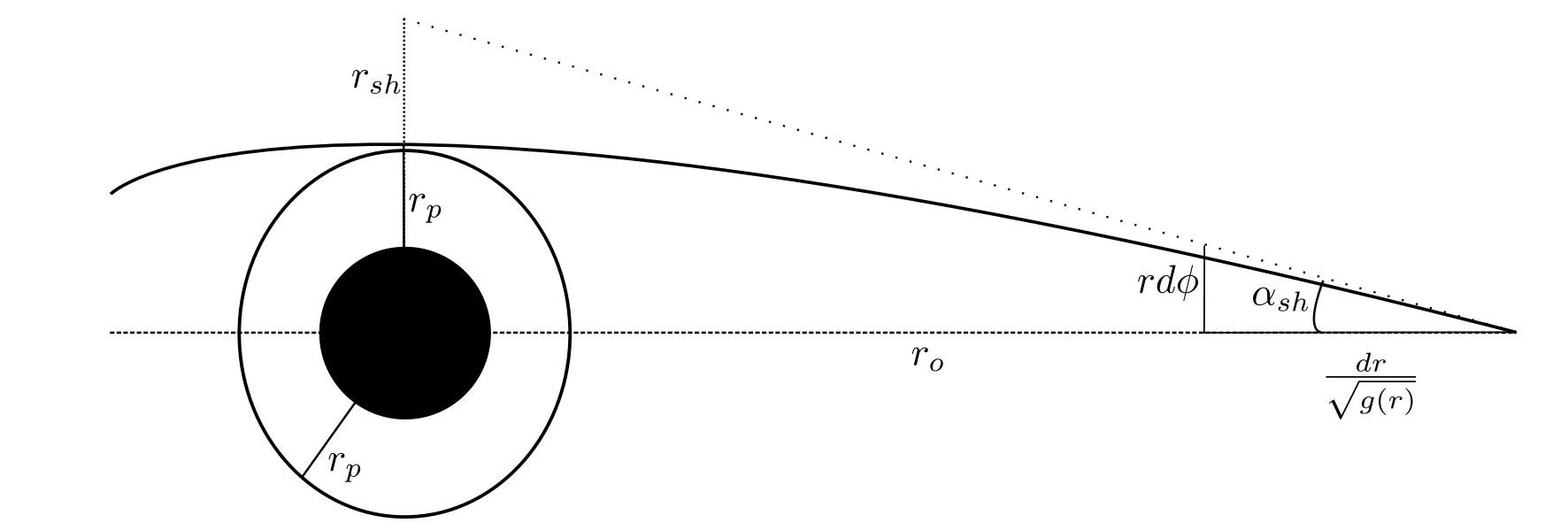}
\caption{The cone of avoidance and the black hole shadow.}
\label{fig:ACOA}
\end{figure}

Now, to define the black hole shadow one must consider the idea of a \textit{cone of avoidance}~\cite{chandrasekhar1998mathematical}(See Fig.\,\ref{fig:ACOA}). At any point on the spacetime manifold, this is defined  as an imaginary cone, such that null rays included in the cone will always cross the black hole event horizon and therefore get trapped forever. The photon which is directed just outside the cone, from the observer's location, will follow a curved trajectory and graze the photon sphere, eventually escaping to the background (Fig.\,\ref{fig:ACOA}). The size of the black hole shadow will be directly related to the angular size of the cone of avoidance at that spacetime point. If there is no source between the black hole and the observer, the observer will see a dark region inside the cone of avoidance and outside it the observer will see the photons coming from the background light source.

The cone of avoidance is generated by the element of proper length along radial and angular directions. Thus, we may define the angular size of the cone as (See Fig.\,\ref{fig:ACOA})
\begin{equation}\label{eq:AoC}
\cot{\alpha_{\text{\tiny{sh.}}}}=\left.\frac{ds_r}{ds_\phi}\right|_{r=r_{\text{\tiny{obs.}}}}=\frac{1}{r\sqrt{g(r)}}\left.\frac{dr}{d\phi}\right|_{r=r_{\text{\tiny{obs.}}}}\;
\end{equation}
We are interested in the null trajectories that are just grazing the $r=r_{\text{\tiny{ph.}}}$ boundary, as defined in Eq.\,(\ref{eq:COC}). Hence, using Eq.\,(\ref{eq:RPHI2}), we may write the above equation with $r=r_{\text{\tiny{ph.}}}$ as
\begin{equation}\label{eq:AR}
\sin{\alpha_{\text{\tiny{sh.}}}}=\frac{r_{\text{\tiny{ph.}}}}{r_{\text{\tiny{obs.}}}}\sqrt{\frac{f(r_{\text{\tiny{obs.}}})}{f(r_{\text{\tiny{ph.}}})}}\;.
\end{equation}
The radius of the shadow will therefore be given by
\begin{equation}\label{eq:RSH}
r_{\text{\tiny{sh.}}}=r_{\text{\tiny{obs.}}}\sin{\alpha_{\text{\tiny{sh.}}}}=r_{\text{\tiny{ph.}}}\sqrt{\frac{f(r_{\text{\tiny{obs.}}})}{f(r_{\text{\tiny{ph.}}})}}\;.
\end{equation}
The black hole shadow angular size and radius are seen to only depend on the $g_{tt}$ component of the metric, unlike the light ray trajectories (also see Appendix A).

Again, for Schwarzschild spacetime, using Eqs.\,(\ref{eq:SF}) and (\ref{eq:SGUO}), the above expressions give the well-know result for the shadow of a Schwarzschild black hole
\begin{equation}\label{eq:RSHSC}
r^{\text{\tiny{Sch.}}}_{\text{\tiny{sh.}}}=3\sqrt{3}M\;.
\end{equation}

The above derivations may alternatively be performed in a Hamiltonian formulation. This approach will be more amenable to the inclusion of the DM plasma effects in the next section. The action of a relativistic particle is
\begin{equation}\label{eq:A1}
S=\int d\zeta \; \frac{1}{2}g_{\mu\nu}U^{\mu}U^{\nu}\;,
\end{equation}
where $\zeta$ is the affine parameter and the four-velocities are defined as $U^{\mu}=dx^\mu/d\zeta$. The covariant Lagrangian may then be identified as 
\begin{equation}\label{eq:CL}
L=\frac{1}{2}g_{\mu\nu}U^{\mu}U^{\nu}\; .
 \end{equation}
Constructing the conjugate momenta, the corresponding Hamiltonian comes out to be 
\begin{equation}\label{eq:H}
 H=\frac{1}{2}g^{\mu\nu}p_{\mu}p_{\nu}\; .
 \end{equation}
 For the case of the photon, which is of pertinent interest, the null geodesic trajectories imply
\begin{equation}\label{eq:PH}
H_{\gamma}=\frac{1}{2}g^{\mu\nu}p_{\mu}p_{\nu}=-\frac{p_t^2}{f(r)}+g(r)p_r^2+\frac{p_\phi^2}{r^2}=0\;.
 \end{equation} 
 Generalised coordinates $t$ and $\phi$ are seen to be cyclic and lead to conservation of the conjugate momenta $p_t$ and $p_\phi$. This, of course, just corresponds to the integrals of motion in Eq.\,(\ref{eq:CQ}). From the Hamilton's equation $\dot{r}=\frac{\partial H}{\partial p_r}$, we get again Eq.\,(\ref{eq:leconst}). The subsequent analysis may be performed as before. The Hamiltonian formulation is more convenient for incorporating the additional dispersion effects from the DM plasma, as we shall see in Sec.\,\ref{sec:mcpshdw} and Appendix B. 
  
%%%%%%%%%%%%%%%%%%%%%%%%%%%%%%%%%%%%%%%

\section{Influence of dark matter plasma on black hole shadows}\label{sec:mcpshdw}
Let us now explore the additional effects due to a DM plasma component, on the propagation of light in the black hole exterior spacetime. We will analyse the case of a static, spherically symmetric black hole embedded in a DM distribution which is also approximately spherically symmetric around it (see Appendix A). We will assume two scenarios for the spherically symmetric DM plasma distribution---constant mass density and radially in-falling.

 The scalar mCP species $\tilde{e}$ and $\tilde{p}$, see  subsection\,\ref{subsec:imps}, will be assumed to constitute a fraction of the total DM thus distributed. The mCP species carrying  a negative charge will be denoted generically by $\tilde{e}$ and the species carrying a positive charge by $\tilde{p}$. We assume that they are not anti-particles of each other. In scenarios where they are, there could be interesting signatures due to DM annihilations in the DM spike regions (see for instance\,\cite{Shelton:2015aqa}). The corresponding charges will be labelled $-\epsilon q_e$ and $+\epsilon q_e$, and the respective masses of each species by $m_{\tilde{e}}$ and $m_{\tilde{p}}$. 
 
 The large number densities in regions of the mCP parameter space  $m_{\text{\tiny{mCP}}}\lesssim \mathcal{O}(1)\,\mathrm{GeV}$, implies that the DM plasma may be treated as a fluid to good approximation for these ranges. We adapt the two-fluid model\,\cite{Synge:1960ueh, breuer1980propagation,breuer1981propagation,perlick2000ray} for plasmas, to the DM plasma case of interest (see Appendix B). Net charge quasi-neutrality will be assumed for the DM plasma, assuming a symmetric fraction for the $\tilde{e}$ and $\tilde{p}$ species\,\cite{Cline:2021itd}. We are primarily interested in the propagation of light rays in the spacetime background with DM plasma. A non-zero pressure in the DM plasma, and a consequent pressure gradient term in the DM plasma force equations (see Appendix B), will not have any effect on the propagation of transverse electromagnetic waves. A non-zero pressure will primarily lead only to modifications of the longitudinal wave dispersion relation; the so called Bohm-Gross relation (see for instance,\,\cite{1998pfp..book.....C,thorne2017modern}). A DM plasma pressure contribution may modify the metric components. In the low-mass mCP regions where it may be important, it will only lead to even larger deviations in the photon sphere and shadow radii, compared to the scenarios we consider. Motivated by these, we will conservatively assume that the DM plasma fluid is cold, pressure-less and non-magnetised. This should help us glean conservative trends on how a DM plasma furnished by mCP particles will modify the photon sphere and black hole shadow radii.
  
 The DM plasma frequency due to the $\tilde{e}$ and $\tilde{p}$ species, at a radial distance `r' is given by (Appendix B)
\begin{equation}\label{eq:mCPAPF}
\tilde{\omega}^2_{p}(r)=\frac{\epsilon^2 q_e^2 }{\epsilon_o\mu}\tilde{N}(r)\;,
\end{equation}
where $\tilde{N}(r)$ is the number density of either mCP species and the reduced mass is given by $\mu=m_{\tilde{e}}m_{\tilde{p}}/(m_{\tilde{e}}+m_{\tilde{p}})$. The corresponding mass density is given by
\begin{equation}
\tilde{\rho}(r)=\left(m_{\tilde{e}}+m_{\tilde{p}}\right)\tilde{N}(r) \; ,
\end{equation}
and the plasma frequency may be equivalently written in terms of this as
\begin{equation}\label{eq:mCPAPF1}
\tilde{\omega}^2_{p}(r)=\frac{\epsilon^2q_e^2 }{\epsilon_om_{\tilde{e}}m_{\tilde{p}}}\tilde{\rho}(r) \; .
\end{equation}

The Hamiltonian for a light ray propagating in a spacetime with cold, pressure-less, non-magnetised plasma may be written\,\cite{perlick2000ray} (see also Appendix B) as
\begin{equation}\label{eq:mCPAPHM}
H_\gamma=\frac{1}{2}\left(g^{\mu\nu}p_{\mu}p_{\nu}+\tilde{\omega}^2_{p}(r)\right)=\frac{1}{2}\left(-\frac{p_t^2}{f(r)}+g(r)p_r^2+\frac{p_\phi^2}{r^2}+\tilde{\omega}^2_{p}(r)\right)=0\;.
\end{equation}
Utilising the constancy of the conjugate momenta corresponding to the cyclic coordinates, the above may be re-written as
\begin{equation}\label{eq:mCPAPH1}
H_{\gamma}=-\frac{E^2}{f(r)}+g(r)p_r^2+\frac{L^2}{r^2}+\tilde{\omega}^2_{p}(r)=0\; .
\end{equation}

From the Hamilton's equations, as motivated in the previous section, the trajectory of the light rays in the present case, with the effects of DM plasma included, may be derived. It is found to be of the form
\begin{equation}\label{eq:mCPAPRPHI}
\frac{dr}{d\phi}=\pm r\sqrt{g(r)}\sqrt{\left[\frac{r^2E^2}{f(r)L^2}\left(1-\frac{\tilde{\omega}^2_{p}(r)f(r)}{E^2}\right)-1\right]} \; .
\end{equation}
Note that the light ray trajectories are not null geodesics anymore, due to the additional electromagnetic effects sourced by the DM plasma.
 
The above equation suggests that for the viable propagation of a photon, one requires
\begin{equation}\label{eq:mCPAPFC0}
\frac{r^2E^2}{f(r)L^2}\left(1-\frac{\tilde{\omega}^2_{p}(r)f(r)}{E^2}\right)-1> 0\; ,
\end{equation}
or equivalently that the gravitationally red-shifted photon frequency must satisfy
\begin{equation}\label{eq:photonpropcond}
 \omega^2(r) \equiv\frac{E^2}{f(r)}>\tilde{\omega}^2_{p}(r)+\frac{L^2}{r^2}\;.
\end{equation}
For $L=0$, this is just the criterion that at a radial distance `r', for photons to propagate through the DM plasma one would require
\begin{equation}\label{eq:opnecessarycond}
\omega^2(r)>\tilde{\omega}_p^2(r)\;.
\end{equation}
This is a familiar result from plasma physics.

Now, Eq.\,(\ref{eq:mCPAPH1}) is of the form
\begin{equation}\label{eq:mCPAPRE}
\frac{1}{2}g(r)p_r^2-\frac{L^2}{2r^2}\left[\frac{E^2r^2}{L^2f(r)}\left(1-\frac{\tilde{\omega}^2_p(r)f(r)}{E^2}\right)-1\right]=0\;,
\end{equation}
which, as in the previous section, may be interpreted to be of the form
\begin{equation}\label{eq:mCPAPRE1}
\frac{1}{2}g(r)p_r^2+\tilde{V}_{eff}=0\;,
\end{equation}
with 
\begin{equation}\label{eq:mCPAPV}
\tilde{V}_{eff}=-\frac{L^2}{2r^2}\left[\frac{E^2r^2}{L^2f(r)}\left(1-\frac{\tilde{\omega}^2_p(r)f(r)}{E^2}\right)-1\right]\;,
\end{equation}
as the relevant effective potential.

Imposing  $\tilde{V}_{eff}=0$ and $\tilde{V'}_{eff}=0$, the modified photon sphere radius may be determined implicitly by the criterion
\begin{equation}\label{eq:mCPAPPSR}
\frac{d}{dr}\left.\left[\frac{r^2}{f(r)}-\frac{\tilde{\omega}^2_p(r)r^2}{E^2}\right]\right|_{r=r_{\text{\tiny{ph.}}}}=0\;.
\end{equation}
With the photon sphere radius hence determined, the altered shadow radius will then be given by
\begin{equation}\label{eq:mCPAPRSH}
r_{\text{\tiny{sh.}}}=r_{\text{\tiny{obs.}}}\sin{\alpha_{\text{\tiny{sh.}}}}=r_{\text{\tiny{ph.}}}\sqrt{\frac{f(r_{\text{\tiny{obs.}}})}{f(r_{\text{\tiny{ph.}}})}\frac{{\left(1-\frac{\tilde{\omega}^2_{p}(r_{\text{\tiny{ph.}}})f(r_{\text{\tiny{ph.}}})}{E^2}\right)}}{{\left(1-\frac{\tilde{\omega}^2_{p}(r_{\text{\tiny{obs.}}})f(r_{\text{\tiny{obs.}}})}{E^2}\right)}}}\;.
\end{equation}
Note in particular that the effect of the DM plasma appears both directly in the shadow radius expression and implicitly through the altered $r_{\text{\tiny{ph.}}}$.

Let us now consider two simple, semi-realistic scenarios for the spherically symmetric DM distribution. The first scenario assumes an approximately constant DM mass density extending over a finite region and the second scenario assumes DM that is radially in-falling in a finite region into the black hole. These simplified cases for the DM distribution around the black hole will help us analytically estimate some of the noteworthy effects and features.

%%%%%%%%%%%%%%
\subsection{Constant Density of Dark Matter Plasma}
\label{sec:cmdsec}

The first simplified case that we will consider is that of a DM distribution with constant mass density, in a spherically symmetric, finite region around the black hole. The DM plasma distribution will be assumed to follow the total DM distribution and will be given by
\begin{equation}\label{eq:mCPAPCDD}
\tilde{\rho}(r)= \begin{cases}
f_{\text{\tiny mCP}}\, \rho_0 \quad & R_{\text{\tiny{Sch.}}}<r<\kappa R_{\text{\tiny{Sch.}}} \;,\\
0 \quad & \kappa R_{\text{\tiny{Sch.}}}\le r \;. \\
\end{cases}
\end{equation}
Here, $R_{\text{\tiny{Sch.}}}$ is the Schwarzschild radius of the central black hole, $\rho_0$ is the constant mass density of the total DM and $\kappa$ is a parameter that specifies the region of validity for the constant mass density approximation. This is a crude approximation that may be motivated by the observed flat DM mass profiles in many galactic cores\,\cite{Salucci:2000ps,Burkert:1995yz}. It is also assumed that during the period of observation that the analysis applies to, the gain in mass by the black hole is negligible compared to its initial mass.

The corresponding mass distribution is
\begin{equation}\label{eq:mCPAPCDMD}
M(r)= \begin{cases}
M_{\text{\tiny{BH}}} \quad & r\le R_{\text{\tiny{Sch.}}} \;,\\
M_{\text{\tiny{BH}}} + \frac{4\pi \rho_0}{3}\left(r^3-R_{\text{\tiny{Sch.}}}^3\right) \quad & R_{\text{\tiny{Sch.}}}<r<\kappa R_{\text{\tiny{Sch.}}} \;, \\
M_{\text{\tiny{BH}}} +\frac{4\pi \rho_0R_{\text{\tiny{Sch.}}}^3}{3}\left(\kappa^3-1\right) \quad & \kappa R_{\text{\tiny{Sch.}}}\le r \; ,\\
\end{cases}
\end{equation}
where $M_{\text{\tiny{BH}}} $ is the mass of the central black hole. The terms proportional to $\rho_0$ are due to the spherically symmetric DM distribution. In our choice of $\kappa$, which determines the finite extent of the DM plasma, we will ensure that $4\pi  \rho_0 R_{\text{\tiny{Sch.}}}^3\left(\kappa^3-1\right)/3 \ll M_{\text{\tiny{BH}}}$, so that the total DM mass contribution is negligible compared to the central black hole mass. This is also motivated by current observations, for instance, of the Sagittarius A* and M87* black holes.

The DM plasma component will effect the erstwhile null geodesics, and through electromagnetic interactions cause a non-geodesic deviation in their paths. The DM plasma frequency correction term in Eq.\,(\ref{eq:mCPAPH1}), relative to Eq.\,(\ref{eq:PH}), is now
\begin{equation}\label{eq:mCPAPCDPF}
\tilde{\omega}^2_p(r)= \begin{cases}
\frac{\epsilon^2q_e^2 f_{\text{\tiny mCP}} \rho_0}{\epsilon_o m_{\tilde{e}}m_{\tilde{p}}} \quad & R_{\text{\tiny{Sch.}}}<r<\kappa R_{\text{\tiny{Sch.}}}\;, \\
0 \quad & \kappa R_{\text{\tiny{Sch.}}}\le r \; .\\
\end{cases}
\end{equation}
 
 The spherically symmetric DM distribution modifies the $g_{tt}$ component of the metric in the following way (see Appendix A)
\begin{equation}\label{eq:mCPAPCDf}
f(r)= \begin{cases}
1-\frac{R_{\text{\tiny{Sch.}}}}{r}-\frac{8\pi G \rho_0}{3c^2r}\left(r^3-R_{\text{\tiny{Sch.}}}^3\right) \quad & R_{\text{\tiny{Sch.}}}<r<\kappa R_{\text{\tiny{Sch.}}} \;,
\vspace{0.1in}\\
1-\frac{R_{\text{\tiny{Sch.}}}}{r}-\frac{8\pi G \rho_0R_{\text{\tiny{Sch.}}}^3}{3rc^2}\left(\kappa^3-1\right) \quad & \kappa R_{\text{\tiny{Sch.}}}\le r \; .
\end{cases}
\end{equation}
Thus, apart from the electromagnetic effect that any mCP component may have on the light ray trajectories, there is also in principle a gravitational effect due to the additional DM mass distribution around the black hole. 

The above point---that there will be both an electromagnetic and gravitational effect due to the DM plasma distribution---motivates us to clearly distinguish these two contributions. Towards demarcating these and clarifying the respective effects in all the subsequent expressions, let us define two, dimensionless quantities in the constant density case
\begin{equation}\label{eq:mCPAPCdp}
\begin{split}
\chi_{d}&\equiv\frac{\epsilon^2q_e^2 f_{\text{\tiny mCP}}\, \rho_0}{\epsilon_om_{\tilde{e}}m_{\tilde{p}} E^2}\;,\\
\chi_{g}&\equiv\frac{8\pi \rho_0GR_{\text{\tiny{Sch.}}}^2}{3c^2}\;.\\
\end{split}
\end{equation}
Note that while $\chi_d$ is solely due to the DM plasma component, the $\chi_g$ term depends on the total DM distribution.

In terms of these dimensionless quantities, we now have in the region of interest $ R_{\text{\tiny{Sch.}}}<r<\kappa R_{\text{\tiny{Sch.}}}$
\begin{eqnarray}\label{eq:pertquantconst}
\tilde{\omega}^2_p(r)&=& \frac{4 \pi^2 c^2\chi_{d} } {\lambda_{\text{\tiny{EM}}}^2} \; ,\nonumber \\
f(r) &=&  1-\frac{R_{\text{\tiny{Sch.}}}}{r}-\chi_{g}\left[\left(\frac{r}{R_{\text{\tiny{Sch.}}}}\right)^2-\left(\frac{R_{\text{\tiny{Sch.}}}}{r}\right)\right]  \; .
\end{eqnarray}
These will be the pertinent quantities through which the DM plasma effects will manifest.

Applying Eq.\,(\ref{eq:pertquantconst}) to Eq.\,(\ref{eq:mCPAPRSH}), we find the formal expression for the modified black hole shadow radius in the present case to be
\begin{equation}\label{eq:mCPAPCDrshe1}
r_{\text{\tiny{sh.}}}=\begin{cases}
	\sqrt{\left(1-\frac{R_{\text{\tiny{Sch.}}}}{r_{\text{\tiny{obs.}}}}-\chi_{g}\frac{R_{\text{\tiny{Sch.}}}}{r_{\text{\tiny{obs.}}}}\left(\kappa^3-1\right)\right)\left(\frac{r_{\text{\tiny{ph.}}}^2}{1-\frac{R_{\text{\tiny{Sch.}}}}{r_{\text{\tiny{ph.}}}}}\right)} ~~~;~~~~~ \kappa<3/2 \;,\\
	\\
	\sqrt{\left(1-\frac{R_{\text{\tiny{Sch.}}}}{r_{\text{\tiny{obs.}}}}-\chi_{g}\frac{R_{\text{\tiny{Sch.}}}}{r_{\text{\tiny{obs.}}}}\left(\kappa^3-1\right)\right)\left(\frac{r_{\text{\tiny{ph.}}}^2}{1-\frac{R_{\text{\tiny{Sch.}}}}{r_{\text{\tiny{ph.}}}}-\chi_{g}\left(\left(\frac{r_{\text{\tiny{ph.}}}}{R_{\text{\tiny{Sch.}}}}\right)^2-\left(\frac{R_{\text{\tiny{Sch.}}}}{r_{\text{\tiny{ph.}}}}\right)\right)}-\chi_{d}r^2_{\text{\tiny{ph.}}}\right)} ~;~\kappa\ge 3/2.
\end{cases}
\end{equation}
Note that the photon sphere radius appearing in the above expression is also modified by the presence of the DM distribution, and it is the modified value that needs to be used above. Also note that due to the presence of the DM plasma dispersion, quantified by $\chi_d$, the shadow radius is now dependent on the electromagnetic wavelength being used for observation.

With the relevant expressions in Eq.\,(\ref{eq:pertquantconst}), we get from Eq.\,(\ref{eq:mCPAPPSR}), implicitly for $r_{\text{\tiny{ph.}}}$
\begin{equation}\label{eq:phradcd}
\begin{split}
	\frac{d}{dr}\left.\left[\frac{r^2}{1-\frac{R_{\text{\tiny{Sch.}}}}{r}-\chi_{g}\frac{R_{\text{\tiny{Sch.}}}}{r_{\text{\tiny{obs.}}}}\left(\kappa^3-1\right)}\right]\right|_{r=r_{\text{\tiny{ph.}}}}=0 ~;~~\kappa<3/2 \;,\\
	\frac{d}{dr}\left.\left[\frac{r^2}{1-\frac{R_{\text{\tiny{Sch.}}}}{r}-\chi_{g}\left(\left(\frac{r}{R_{\text{\tiny{Sch.}}}}\right)^2-\left(\frac{R_{\text{\tiny{Sch.}}}}{r}\right)\right)}-\chi_{d}\,r^2\right]\right|_{r=r_{\text{\tiny{ph.}}}}=0 ~;~~\kappa\ge 3/2 \;.\\
\end{split}
\end{equation}
Notice that for $\kappa<3/2$, the effect of the DM plasma cloud is only through its gravitational effect. 

\begin{figure}
 	\center
 	\includegraphics[scale=0.6]{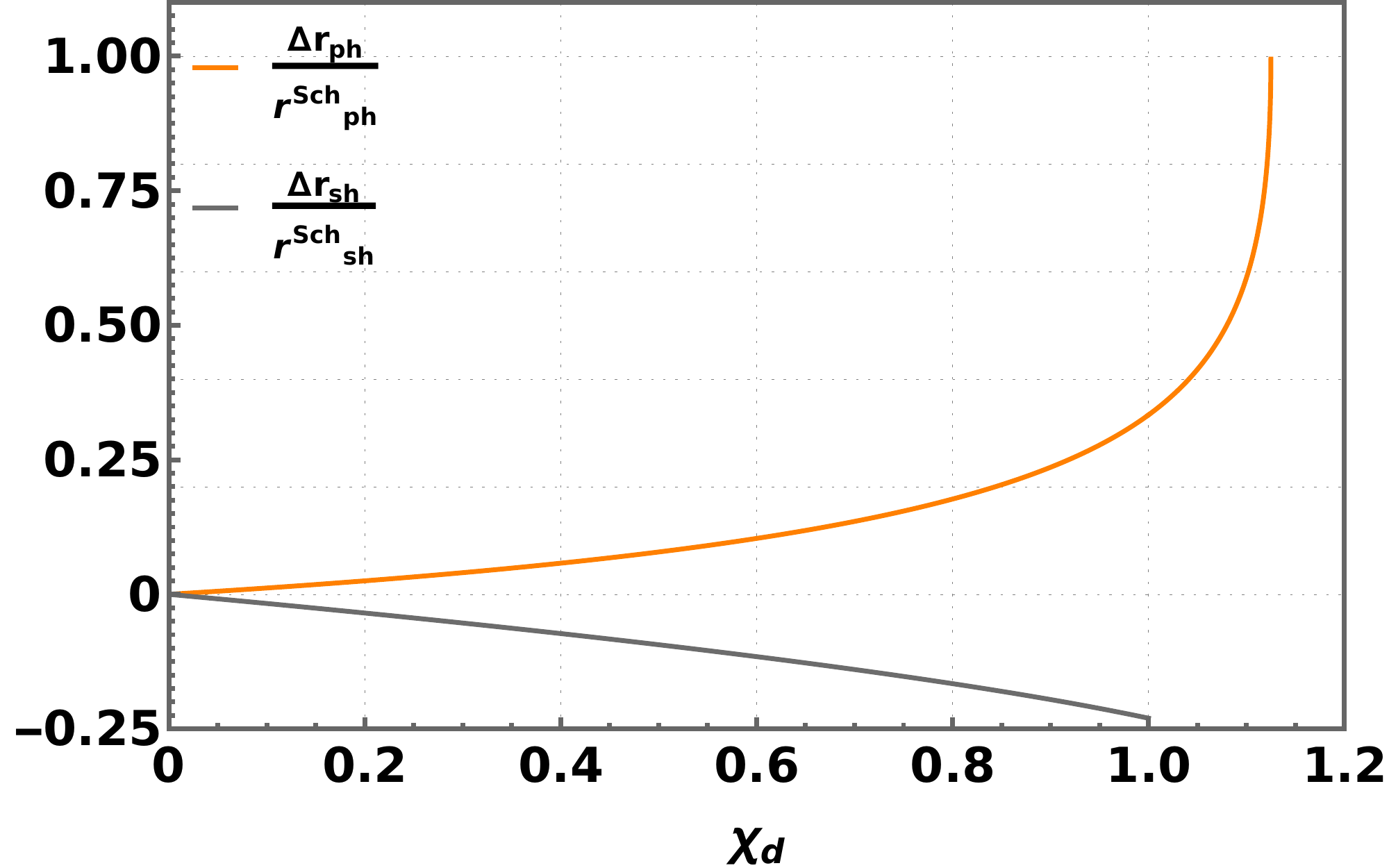}
 	\caption{Variation in the photon sphere and black hole shadow radius as a function of $\chi_{d}$, assuming constant DM density. Here, we have defined ${\Delta r_{\text{(\tiny{ph.,sh.})}}}/{r^{\text{\tiny{Sch.}}}_{\text{(\tiny{ph.,sh.})}}} =\left({ r_{\text{(\tiny{ph.,sh.})}}-r^{\text{\tiny{Sch.}}}_{\text{(\tiny{ph.,sh.})}}}\right)/{r^{\text{\tiny{Sch.}}}_{\text{(\tiny{ph.,sh.})}}}$. It is noted that the photon sphere radius increases as a function of $\chi_{d}$. At the threshold value $\chi_{d}=1$ there is sudden increase in photon sphere radius, as discussed in the text, and beyond $\chi_{d}\ge9/8$, the photon sphere is non-existent. Remarkably, in contrast, the shadow radius actually decreases as $\chi_{d}$ increases, due to the dispersion of the null rays. The shadow radius could changes as much as $\sim 25\%$, and for $\chi_{d}\gtrsim1$ the DM plasma completely attenuates the electromagnetic radiation.}
 	\label{fig:CMDchid}
 \end{figure}

The above equations may be solved numerically and we show the results in Fig.\,\ref{fig:CMDchid}. One observes that the photon sphere radius increases with increasing $\chi_d$ and the shadow radius decreases with increasing $\chi_d$. The functional dependences also display unique characteristics. When we discuss the case of radially in-falling DM plasma we will observe that the general trends are similar, but the functional dependences and thresholds are very different.
%%%%%
 %%%%%

To get an analytic understanding of the contributions and their interplay, we may also perturbatively solve Eq.\,(\ref{eq:phradcd}), in the regime where $\chi_{g,d} \ll 1$, and give an approximate analytic expression for the shadow radius from Eq.\,(\ref{eq:mCPAPCDrshe1}). However, note that this regime is not valid for some of the mCP parameter space regions we analyse, and exact expressions are used in the full analyses we perform. 

In the regions where it is a good approximation, the above procedure gives an approximate analytic expression for the modified photon sphere radius. To leading order in $\chi_{g,d}$, it is of the form
\begin{equation}\label{eq:mCPAPCDrp}
\frac{r_{\text{\tiny{ph.}}}~}{R_{\text{\tiny{Sch.}}}}\simeq\frac{3}{2}+\frac{\chi_{d} }{6}-\frac{3 \chi_{g} }{2}+\frac{5 \chi_{d} ^2}{54}-\frac{29 \chi_{d}  \chi_{g}
}{12}+O(\chi_{d,g}^3) \; .
\end{equation}

Similarly, for $\chi_{g,d} \ll 1$, on substituting the above in Eq.\,(\ref{eq:mCPAPCDrshe1}), we have for the black hole shadow radius, to leading order, the expression 
\begin{eqnarray}\label{eq:mCPAPCDrsh}
\frac{r_{\text{\tiny{sh.}}}}{R_{\text{\tiny{Sch.}}}}&\simeq& \frac{3 \sqrt{3}}{2}\left(1-\frac{R_{\text{\tiny{Sch.}}}}{r_{\text{\tiny{obs.}}}}\right)^{\frac{1}{2}}-\frac{\sqrt{3} \chi_{d} }{4}\left(1-\frac{R_{\text{\tiny{Sch.}}}}{r_{\text{\tiny{obs.}}}}\right)^{\frac{1}{2}} \nonumber \\
&+&\frac{3 \sqrt{3} \chi_{g}}{16}\left(-15-4\kappa^3+19\frac{r_{\text{\tiny{obs.}}}}{R_{\text{\tiny{Sch.}}}}\right)\left(\frac{R_{\text{\tiny{Sch.}}}^2}{r_{\text{\tiny{obs.}}}(r_{\text{\tiny{obs.}}}-R_{\text{\tiny{Sch.}}})}\right)^{1/2}+O(\chi_{d,g}^2)\; .
\end{eqnarray}

To get a sense for the relative magnitudes of the $\chi_{g,d}$ contributions in Eqs.\,(\ref{eq:mCPAPCDrp}) and (\ref{eq:mCPAPCDrsh}), let us estimate them with some reference parameter values. This gives
\begin{eqnarray}\label{eq:mCPAPCdp1}
\chi_{d}&\equiv&\frac{\epsilon^2q_e^2f_{\text{\tiny mCP}} \rho_0}{\eta\epsilon_om^2_{\tilde{p }} \left(\frac{2\pi c}{\lambda_{\text{\tiny{EM}}}}\right)^2}=1.74\times10^{-8} \times  \frac{\epsilon^2}{\eta} \times \left(\frac{eV/c^2}{ m_{\tilde{p}}}\right)^2\times  \left(\frac{f_{\text{\tiny mCP}}\rho_0}{ \solarmass  /(Kpc)^3}\right)\times \left(\frac{\lambda_{\text{\tiny{EM}}}}{ mm }\right)^2 \;,\nonumber \\
\chi_{g}&\equiv&\frac{8\pi  \rho_0GR_{\text{\tiny{Sch.}}}^2}{3c^2}= 3.67\times10^{-45}\times\left(\frac{\rho_0}{ \solarmass  /(Kpc)^3}\right)\times\left(\frac{M_{\text{\tiny{BH}}} }{\solarmass  }\right)^2 \; .
\end{eqnarray}
Here, $\eta=m_{\tilde{e}}/m_{\tilde{p}}$ and $\lambda_{\text{\tiny{EM}}}$ is the wavelength of the electromagnetic radiation. As quantified explicitly above, $\chi_d$ depends on the electromagnetic wavelength. Thus, as remarked under Eq.\,(\ref{eq:mCPAPCDrshe1}), due to the dependence of the black hole shadow radius on $\chi_d$, it too will now depend on the electromagnetic wavelength being utilised for observation.

Based on Eq.\,(\ref{eq:mCPAPCdp1}), we conclude that the gravitational effect of the DM plasma will be very small. The gravitational effect due to the total DM distribution is also mostly negligible. For example, even taking a substantial---but reasonable and motivated\,\cite{Gondolo_1999,Lacroix2015}---value for the DM plasma mass density $f_{\text{\tiny mCP}}\rho_0=6.9\times10^{13} \solarmass/(Kpc)^3$, with $f_{\text{\tiny mCP}}=0.1\%$, and black hole mass $M_{\text{\tiny BH}}=6.5\times10^9 \solarmass$, one obtains for the total DM gravitational contribution $\chi_{g}\sim 10^{-11}$. On the other hand, the typical values of DM plasma $\chi_d$ are much larger. For example, taking $\eta=1$, $\lambda_{\text{\tiny{EM}}}=1.3\,\mathrm{mm}$, $\epsilon=10^{-17}$ and $m_{\text{\tiny mCP}}=10^{-13}$ eV, we estimate a value $\chi_d\sim0.02$. In contrast, for ordinary Hydrogen plasma $\chi_{d}$ would come out to be $4.2\times10^{-9}$, with any consequent effect on the shadow radius that is orders of magnitude smaller, relative to DM plasmas. For a fixed value of $\epsilon$, one also notes that the effects are diminished for increasing values of $m_{\tilde{p}}$. For $\epsilon\lesssim 10^{-14}$, as motivated in Sec.\,\ref{sec:mcps}, the most interesting effects are in the sub-eV mCP mass ranges.

Based on these observation, we may effectively take $\chi_{g}=0$ and obtain a near exact solution for the photon sphere radius, in terms of $\chi_{d}$ alone, in the constant mass density case. One obtains for the photon sphere radius with $\chi_{g}=0$
\begin{equation}\label{eq:mCPAPCDpha}
\left.\frac{r_{\text{\tiny{ph.}}}~}{R_{\text{\tiny{Sch.}}}}\right|_{\chi_{g}=0}=\begin{cases}
\frac{3+\sqrt{9-8\chi_{d}}-4\chi_{d}}{4(1-\chi_{d})}  \hquad &\chi_{d} \ne 1\; ,\\
2  \hquad \hquad &\chi_{d} = 1 \; .
\end{cases} 
\end{equation}

Similarly, in this case, an analytic expression for the black hole radius may also be obtained of the form
\begin{equation}\label{eq:mCPAPCDsha}
\left.\frac{r_{\text{\tiny{sh.}}}~}{R_{\text{\tiny{Sch.}}}}\right|_{\chi_{g}=0}=\frac{1}{4}\sqrt{\frac{-18(3+\sqrt{9-8\chi_{d}})+8(9+2\sqrt{9-8\chi_{d}}-2\chi_{d})\chi_{d}}{\chi_{d}-1}}\left(\sqrt{1-\frac{R_{\text{\tiny{Sch.}}}}{r_{\text{\tiny{obs.}}}}}\right) ,~\chi_{d}  < 1\,.
\end{equation}

Now, one point to note is that for transmission through the DM plasma we require from Eq.\,(\ref{eq:opnecessarycond}), at a specific point in spacetime, the necessary condition
\begin{equation}
\tilde{\omega}^2_p(r)< \omega^2(r)\;.
\end{equation}
This criteria translates to
\begin{equation}
 \chi_{d}<\frac{1}{1-\frac{R_{\text{\tiny{Sch.}}}}{r}-\chi_{g}\left(\left(\frac{r}{R_{\text{\tiny{Sch.}}}}\right)^2-\left(\frac{R_{\text{\tiny{Sch.}}}}{r}\right)\right)}\;.
 \end{equation}
 
  %%%%%
   \begin{figure}
   		\center
   		\subfloat[]
   		{
   			\includegraphics[scale=0.175]{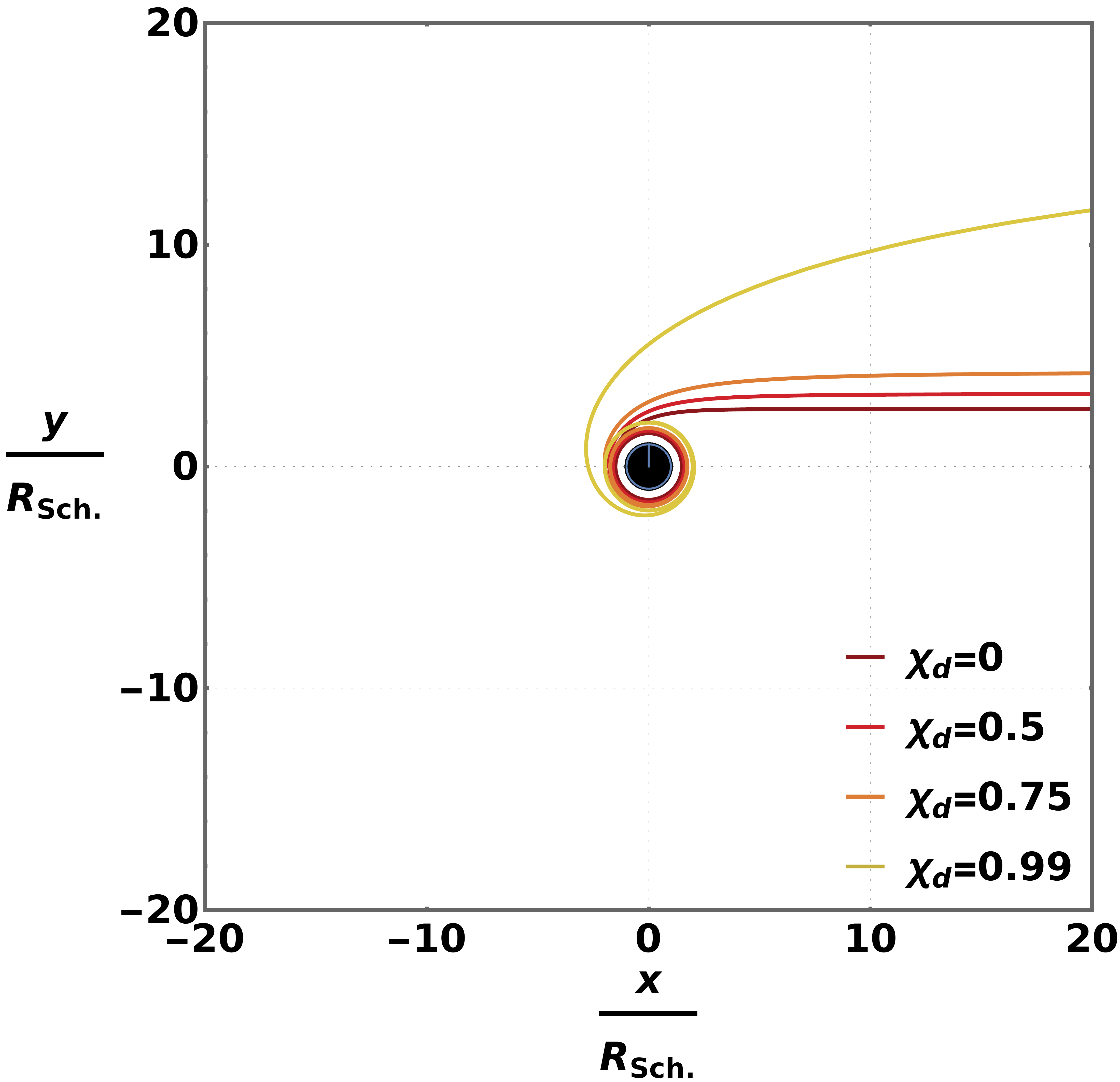}}
   		\subfloat[]
   		{\includegraphics[scale=0.192]{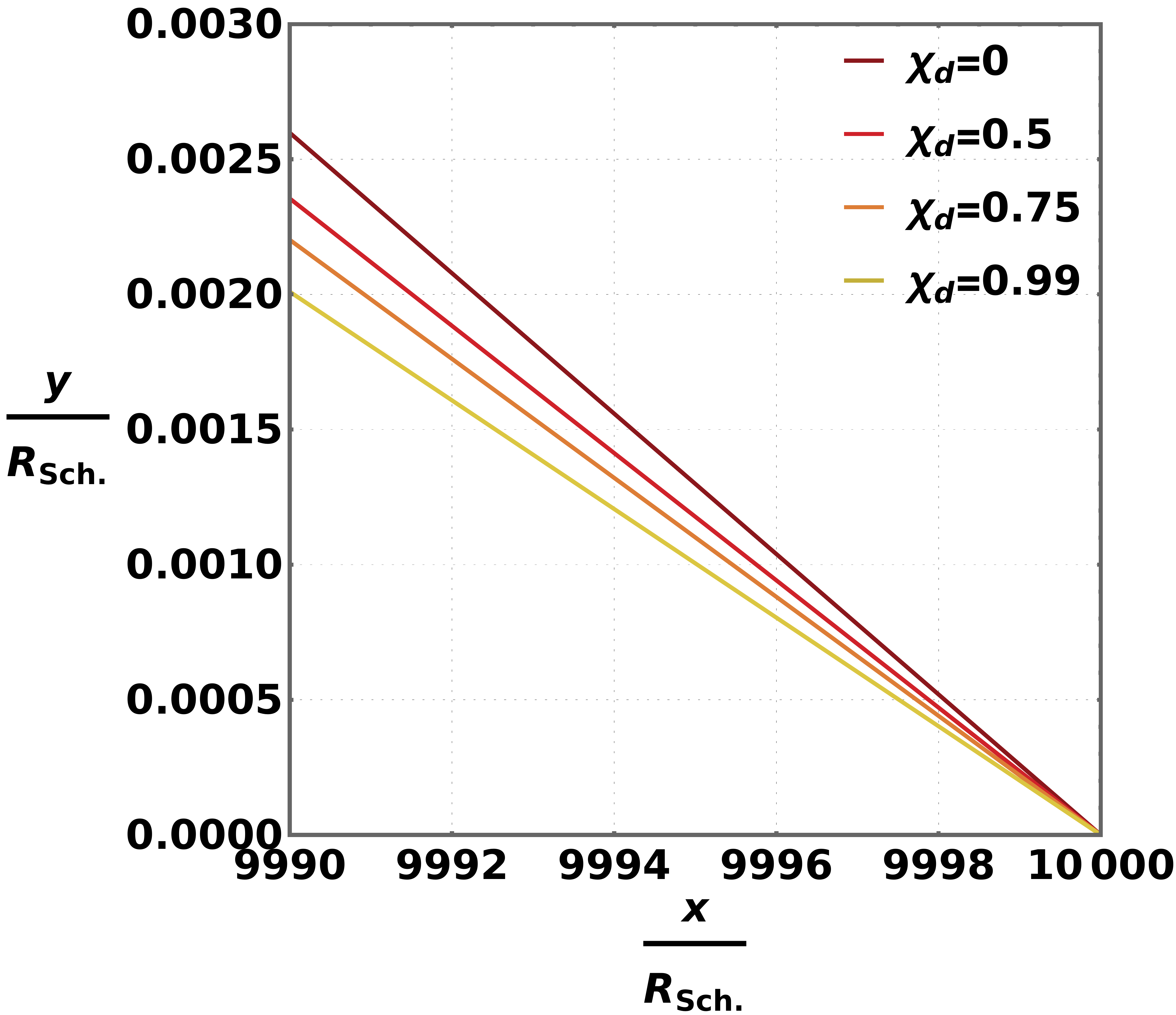}	}\\
   		\subfloat[]
   		{\includegraphics[scale=0.275]{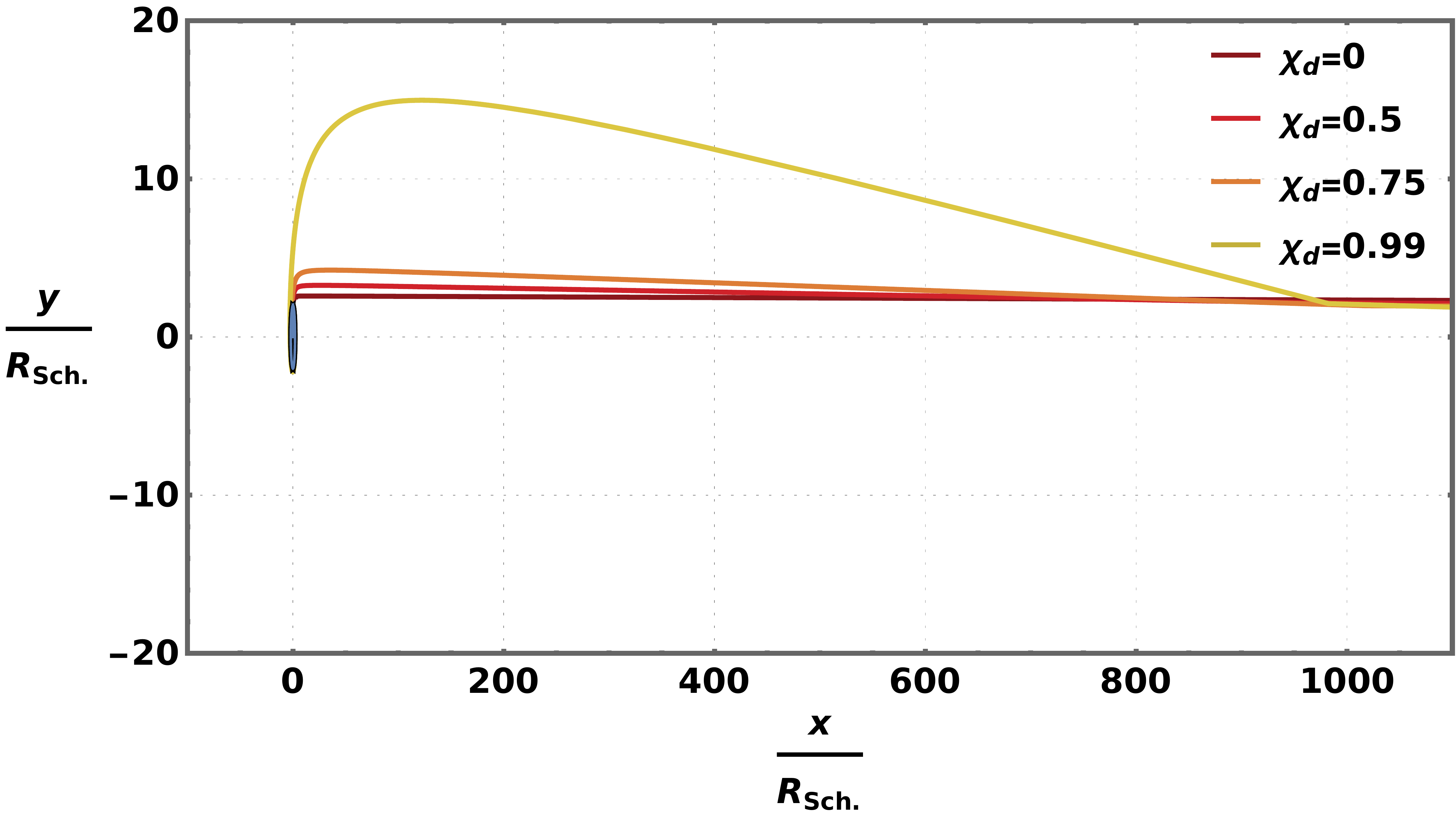}	}
   	\caption{The trajectory of the outgoing light rays (i.e. towards the observer) are shown for the case of constant DM density. For clarity, we are separately focusing on three distinct regions in each subfigure---(a) In the very-near vicinity of the Schwarzschild blackhole (b) From the perspective of the observer, thus illustrating the cone of avoidance in each case (c) In the local neighbourhood of the Schwarzschild black hole, showing how the null trajectories globally approach the black hole. The figures are depicting the equatorial plane of the black hole. In the figures, $(x/R_{\text{\tiny{Sch.}}},y/R_{\text{\tiny{Sch.}}})=(0,0)$ is the centre of the black hole and observer location is being taken as $(x/R_{\text{\tiny{Sch.}}},y/R_{\text{\tiny{Sch.}}})=(10^4,0)$. In the figures, we have taken $f_{\text{\tiny mCP}} \rho_0=6.9\times10^{13} \solarmass/(Kpc)^3$, $M_{\text{\tiny BH}}=6.5\times10^9 \solarmass$, $m_{\tilde{e}}/m_{\tilde{p}}=1$ and $\lambda=1.3\,\mathrm{mm}$. We have assumed a value $\kappa=10^3$, corresponding to a width of a few parsecs, in accordance with expectations from DM spike studies\,\cite{Gondolo_1999,Lacroix2015}. Distinct outgoing trajectories are shown for different $\chi_{d}$. The light ray trajectories in (a) show that the photon sphere increases as $\chi_{d}$ increases. Counterintuitively, the trajectories in (b) from the perspective of the observer show that the shadow radius nevertheless decreases as $\chi_{d}$ increases. This as we explain in the text is due to the refractive nature of the surrounding DM plasma and the effect at the interfaces.}
   	\label{fig:CMDlrt}
   \end{figure}
    %%%%%% 
 As $\chi_d$ is a constant in the present case, we have the necessary criterion
 \begin{equation}\label{eq:cmdemtrans}
 \chi_{d}\lesssim\frac{\kappa}{\kappa-1} \; ,
 \end{equation}
 for non-attenuated transmission through the DM plasma distribution. When $\chi_d$ increases beyond this value, the DM plasma would completely attenuate the electromagnetic radiation. Note that for any value of $\kappa \gg 1$ we effectively have the necessary condition as $\chi_d \lesssim1$. As we will see below, for $\chi_d > \frac{9}{8}$ there will not be a photon sphere. As already noted from Eqs.\,(\ref{eq:mCPAPCdp}) and \,(\ref{eq:mCPAPCdp1}), $\chi_d$ implicitly depends on the electromagnetic wavelength under consideration. This necessary condition may also therefore be written explicitly as
 \begin{equation}\label{eq:cmdemtranswave}
 \lambda_{\text{\tiny{EM}}} < \left(\frac{4\pi^2c^2\epsilon_om_{\tilde{e}}m_{\tilde{p}} \kappa}{\epsilon^2q_e^2 f_{\text{\tiny mCP}}\, \rho_0(\kappa-1)}\right)^\frac{1}{2}\;.
 \end{equation}
 For instance, taking $f_{\text{\tiny mCP}}\, \rho_0 =6.9\times10^{13}\, \solarmass/(Kpc)^3$, $m_{\tilde{e}}=m_{\tilde{p}}$, $\kappa=10^3$ and adopting $\epsilon=10^{-17}$, and $m_{\tilde{p}}= 10^{-15}$ eV, we obtain $\lambda_{\text{\tiny{EM}}}<0.091$ mm. Similarly for $\epsilon=10^{-17}$, and $m_{\tilde{p}}=10^{-13}$ eV, we obtain $\lambda_{\text{\tiny{EM}}}<9.1$ mm.
 
  Let us now discuss our key findings. As we have already seen, since $\chi_{g}\rightarrow 0$ the gravitational effect of the DM plasma will be insignificant. The dispersion effect of the DM plasma as quantified by $\chi_d$ will be the dominant effect. Furthermore, for $\chi_{d}\ll1$, Eq.\,(\ref{eq:mCPAPCDrp}) and Eq.\,(\ref{eq:mCPAPCDrsh}) implies that the photon sphere and shadow radius will be close to 3/2 $R_{\text{\tiny{Sch.}}}$ and ${3\sqrt{3}}/{2}~ R_{\text{\tiny{Sch.}}}$. 

In Fig.\,\ref{fig:CMDchid} we show how the photon sphere and shadow radii vary for different values of the DM plasma dispersion parameter $\chi_{d}$. For computing and plotting these, we have used the exact expressions. The presence of the DM plasma increases the radial position of the maxima of the effective potential in Eq.\,(\ref{eq:mCPAPV}). Therefore, the radius of the photon sphere would increase as $\chi_{d}$ becomes larger. For ordinary baryonic plasma, analogous effects on the photon sphere and shadow radius are constant and moreover negligible\,\cite{PhysRevD.92.104031,li2021gravitational}.

Interestingly, there is a sudden increase in photon sphere radius in the range $1\leq \chi_{d}\leq 9/8$. This is because around $\chi_d \sim 1$, the dispersive effect of the DM plasma starts overcoming the gravitational potential. This may be directly verified from Eq.\,(\ref{eq:mCPAPV}), in which for $\tilde{\omega}^2_p /E^2=\chi_{d}>1$, the DM plasma term starts to dominate. We note that
 \begin{equation}
 \left.{r_{\text{\tiny{ph.}}}}\right|_{\chi_{d}\to9/8}\to3\,{R_{\text{\tiny{Sch.}}}}\, .
 \end{equation}
For $\chi_{d}\ge9/8$, there won't be any photon sphere because the DM plasma cloud alters the effective potential such that it has no maxima. It is observed that the DM plasma cloud can increase the photon sphere radius upto $100\%$. Beyond $\chi_d\sim 1$, for large values of $\kappa$, the DM plasma itself becomes opaque to the electromagnetic radiation.

 An important point to note is that though the DM plasma cloud causes the the photon sphere radius to \textit{increase} as $\chi_d$ takes on larger values, the corresponding change in the dispersion alters the photon trajectory in such a way that the shadow radius actually \textit{decreases} from the perspective of the distant observer. Thus the refractive nature of the DM plasma is acting in opposition to the focusing effect of spacetime. 
 
  From Fig.\,\ref{fig:CMDchid} it is observed that the shadow radius decreases from its value $3\sqrt{3}/{2}\,\mathrm{R}_{\text{\tiny{Sch.}}}$, at $\chi_d=0$, and approaches $2\,\mathrm{R}_{\text{\tiny{Sch.}}}$  as $\chi_{d}$ approaches 1,
 \begin{equation}\label{eq:bhrlimit}
 \left.{r_{\text{\tiny{sh.}}}}\right|_{\chi_{d}\to1}\to2\,{R_{\text{\tiny{Sch.}}}}\, .
  \end{equation} 
 
 For large $\kappa$ and $1\leq \chi_{d}\leq 9/8$, despite the presence of the photon sphere, the shadow boundary of the blackhole may not actually be visible to the distant observer. As mentioned earlier, for $\chi_{d}\gtrsim1$ the plasma attenuates the transmission of EM radiation and the DM plasma becomes opaque.
 
 Note also that that as $\chi_{d}\to1$
 \begin{equation}
 \left.{r_{\text{\tiny{ph.}}}}\right|_{\chi_{d}\to1}\to2\,{R_{\text{\tiny{Sch.}}}}\, .
 \end{equation} 
 Comparing this result to the expression in Eq.\,(\ref{eq:bhrlimit}), we see remarkably that in this limit the photon sphere and shadow radius are almost coincident. This is a consequence of the DM plasma completely canceling out the focusing effect of the spacetime. 
 
  All the above observations may also be deduced by explicitly looking at the null ray trajectories in the spacetime where there is DM plasma present. These null ray trajectories are shown in Fig.\,\ref{fig:CMDlrt}. We have assumed the value $f_{\text{\tiny mCP}}=0.1\%$, consistent with Eq.\,(\ref{eq:bcc}). Furthermore, we have taken $f_{\text{\tiny mCP}}\rho_0=6.9\times10^{13} \solarmass/(Kpc)^3$ $\simeq 2.6\times10^6 \, \mathrm{GeV}\mathrm{/cm}^3$, which is a conservative value considering what dark matter spike profiles suggest\,\cite{Lacroix2015} in many cases. The black hole mass has been assumed to be $M_{\text{\tiny BH}}=6.5\times10^9 \solarmass$ and the electromagnetic radiation wavelength as $\lambda_{\text{\tiny{EM}}}=1.3\,\mathrm{mm}$\,\cite{Akiyama:2019cqa}, in the radio-wave range. We have also taken $\kappa=10^3$, which for the above $M_{\text{\tiny BH}}$ corresponds to a DM plasma extent of a few parsecs. This is consistent with expectations from DM spike studies\,\cite{Gondolo_1999,Lacroix2015}. $\eta\equiv m_{\tilde{e}}/m_{\tilde{p}}=1$ is taken for reference. In these plots, the outgoing light ray trajectories from the black hole to the observer are shown, for different value of $\chi_{d}$, assuming a constant DM mass density. The incoming light rays to the black hole are assumed to originate from a background electromagnetic source, both of which are not shown in the figure. From subfigure (a) in Fig.\,\ref{fig:CMDlrt} we clearly see that the photon sphere radius does increase with $\chi_d$. From subfigure (b) in Fig.\,\ref{fig:CMDlrt} we note that the black hole shadow radius in contrast decreases on increasing $\chi_d$.

 Let us turn to explicitly quantifying some of the interesting regions in the $\epsilon-m_{\text{\tiny mCP}}$ parameter space. Again, let us for reference take $f_{\text{\tiny mCP}}=0.1\%$, $f_{\text{\tiny mCP}}\rho_0=6.9\times10^{13} \solarmass/(Kpc)^3$, $M_{\text{\tiny BH}}=6.5\times10^9 \solarmass,$ and $\lambda_{\text{\tiny{EM}}}=1.3\,\mathrm{mm}$. In most regions of the parameter space the effects on the shadow radius is negligible. For instance, for $\epsilon=10^{-17}$ and $m_{\text{\tiny mCP.}}>10^{-12}$ eV the change in photon sphere and shadow radii are negligible (see Table \ref{tab:mCPAPCDem}). The results of a few representative points where the effects approach potentially observable values are also shown in Table \ref{tab:mCPAPCDem}. From the table we see that, for example, at  $\epsilon=10^{-17}$ and $m_{\text{\tiny mCP}}= 1.5 \times10^{-14}\,\mathrm{eV}$ relative change in photon sphere radius may be as high as $24\%$, and that in the shadow radius as large as $20\%$. The maximum change in shadow radius in the viable mCP parameter space regions is found to be around $25\%$. Furthermore, we also note that there is a sudden increase in photon sphere and shadow radii for $\epsilon \sim 10^{-17}$ and $m_{\text{\tiny mCP}}\sim 10^{-14}$. This, as already detailed earlier, is due to the refractive effect of plasma becoming significant as $\chi_d\sim O(1)$. In certain region of the allowed $\epsilon-m_{\text{\tiny mCP}}$ parameter space the radio waves are completely attenuated as the DM plasma becomes opaque to them. This happens around where the necessary condition $\tilde{\omega}^2_p(r)< \omega^2(r)$ is no longer satisfied.
 
 It is observed that the shadow radius decreases as $\epsilon$ increases or $m_{\text{\tiny mCP}}$ decreases, but the photon sphere radius follows an opposite trend i.e. it increases with increasing $\epsilon$ or decreasing $m_{\text{\tiny mCP}}$. This is explained by the fact that the shadow radius is determined by the cone of avoidance from the perspective of the distant observer, and the DM plasma in the intervening space has a refractive effect as well on the light ray trajectories.

 In Table\,\ref{tab:mCPAsheff} we show the estimates for a few galactic black holes, and representative mCP parameters. We are neglecting the spins of these black holes. The astrophysical parameters corresponding to the DM distributions are only very crude estimates, based on available literature (see for instance\,\cite{2015Lacroix,2018Lacroix,2020Fortes}). 
 
 Since we are discussing the slightly artificial case of a constant DM mass density, where we assume the DM plasma extends only till a certain boundary (quantified by $\kappa\,\mathrm{R}_{\text{\tiny{Sch.}}}$), there is also an effect at the interface that refracts the light ray trajectories. When we consider the slightly more realistic case of in-falling DM plasma in the next subsection we will nevertheless see that the main attributes are still preserved. Thus, the broader observations and semi-quantitative features are robust and not mere artefacts of the astrophysical model scenarios. 
 %%%%%%
   \begin{table}[H]
   	\center
   	\renewcommand{\arraystretch}{1.25}
   	\begin{tabular}{|c|c|c|c|c|c|c| } 
   		\hline 
   		~$\epsilon$~&~$m_{\text{\tiny mCP.}}~(eV)$~&~$\chi_{d}$~& $r_{\text{\tiny{ph.}}}~ (R_{\text{\tiny{Sch.}}})$~& ~$\frac{\Delta r_{\text{\tiny{ph.}}}}{r^{\text{\tiny{Sch.}}}_{\text{\tiny{ph.}}}} $~& $r_{\text{\tiny{sh.}}}~ (R_{\text{\tiny{Sch.}}})$~& ~$\frac{\Delta r_{\text{\tiny{sh.}}}}{r^{\text{\tiny{Sch.}}}_{\text{\tiny{sh.}}}}$~\\
   		\hline
   		\multirow{1}{*}{$10^{-14}$}&~$10^{-10}$~&~0.02~&~$1.503$~&~$0.23\%$~&~$2.589$~&~$-0.34\%$~\\
   		\hline
   		\multirow{3}{*}{$10^{-15}$}&~$10^{-10}$~&~$\sim$ 0~&~$1.500$~&~$\sim 0\%$~&~$2.597$~&~$\sim 0\%$~\\
   		&~$10^{-11}$~&~0.02~&~$1.503$~&~$ 0.23\%$~&~$2.589$~&~$ -0.34\%$~\\
   		\hline
   		\multirow{5}{*}{$10^{-16}$}&~$10^{-10}$~&~$\sim$ 0~&~$1.500$~&~$\sim 0\%$~&~$2.598$~&~$\sim 0\%$~\\
   		&~$10^{-11}$~&~$\sim$ 0~&~$1.500$~&~$\sim 0\%$~&~$2.598$~&~$\sim 0\%$~\\
   		&~$10^{-12}$~&~0.02~&~$1.503$~&~$ 0.23\%$~&~$2.589$~&~$- 0.34\%$~\\
   		&~$5\times10^{-13}$~&~0.08~&~$1.514$~&~$0.94\%$~&~$2.562$~&~$-1.4\%$~\\
   		&~$1.5\times10^{-13}$~&~0.90~&~$1.856$~&~$24\%$~&~$2.089$~&~$-20\%$~\\
   		\hline
   		\multirow{6}{*}{$10^{-17}$}&~$10^{-11}$~&~$\sim$ 0~&~$1.500$~&~$\sim 0\%$~&~$2.598$~&~$\sim 0\%$~\\
   		&~$10^{-12}$~&~$\sim$ 0~&~$1.500$~&~$\sim 0\%$~&~$2.598$~&~$\sim 0\%$~\\
   		&~$10^{-13}$~&~0.02~&~$1.503$~&~$ 0.23\%$~&~$2.589$~&~$- 0.34\%$~\\
   		&~$5\times10^{-14}$~&~0.08~&~$1.514$~&~$0.94\%$~&~$2.562$~&~$-1.4\%$~\\
   		&~$1.5\times10^{-14}$~&~0.90~&~$1.856$~&~$24\%$~&~$2.089$~&~$-20\%$~\\
   		&~$1\times10^{-14}$~&~2.03~&~(No photon sphere)~&~$-$~&~(No shadow)~&~$-$~\\
   		\hline
   	\end{tabular}
 	\caption{Photon sphere and shadow radii are shown for a few representative points in the viable mCP parameter space, for the constant mass density case. We have taken $f_{\text{\tiny mCP}} \rho_0=6.9\times10^{13} \solarmass/(Kpc)^3$, $M_{\text{\tiny BH}}=6.5\times10^9 \solarmass$, $\eta=1$ and $\lambda=1.3\,\mathrm{mm}$. We also take $\kappa=10^3$, which for the above $M_{\text{\tiny BH}}$ would correspond to a width of a few parsecs\,\cite{Gondolo_1999,Lacroix2015}. The observer is assumed to be very far, relative to any other relevant length scales. We list the relative changes in the photon sphere and shadow radii, ${\Delta r_{\text{(\tiny{ph.,sh.})}}}/{r^{\text{\tiny{Sch.}}}_{\text{(\tiny{ph.,sh.})}}} =\left({ r_{\text{(\tiny{ph.,sh.})}}-r^{\text{\tiny{Sch.}}}_{\text{(\tiny{ph.,sh.})}}}\right)/{r^{\text{\tiny{Sch.}}}_{\text{(\tiny{ph.,sh.})}}}$. We note that in general for higher $\epsilon$ and lower $m_{\text{\tiny mCP}}$ values, the change in photon sphere and shadow radii are significant. In other regions, the variations are in general insignificant and in all likelihood unobservable. Interestingly, in certain neighbourhoods there is a sudden increase in photon sphere and shadow radii; for example near $\epsilon \sim 10^{-17}$ and $m_{\text{\tiny mCP}}\sim 10^{-14}$. As explained in the text, this is because in these regions $\chi_{d}\sim O(1)$ and the dispersive effect from the DM plasma overwhelms the gravitational potential term.}
   	\label{tab:mCPAPCDem}
   \end{table}
   %%%%%%
 %%%%%%   
 \begin{table}[H]
 	\center
 	\renewcommand{\arraystretch}{1.25}
 	\begin{tabular}{|c|c|c|c|c|c|c|c| } 
 		\hline 
 		Blackhole&Mass $(\solarmass)$&$f_{\text{\tiny  mCP}}\rho_o \,\mathrm{(GeV/cm^3)}$ &$\epsilon$&$m_{\text{\tiny mCP}}\,\mathrm{(eV)}$&$\chi_{d}$& $\frac{\Delta r_{\text{\tiny{ph.}}}}{r^{\text{\tiny{Sch.}}}_{\text{\tiny{ph.}}}} $& $\frac{\Delta r_{\text{\tiny{sh.}}}}{r^{\text{\tiny{Sch.}}}_{\text{\tiny{sh.}}}}$\\
 		\hline
 		\multirow{2}{*} ~M$87$*~&~$6.5 \times 10^{9}$~&~$10^6$~&~$10^{-17} $~&~$2.5\times10^{-14} $~&~0.12~&~$1.5\%$~&~$-2.1\%$~\\
 		~~&~$ $~&~$ $~&~$10^{-18} $~&~$2\times10^{-15} $~&~0.19~&~$2.4\% $~&~$ -3.4\%$~\\
 		\hline
 		\multirow{2}{*} ~Sgr A*~&~$4.3 \times 10^{6}$~&~$10^9$~&~$10^{-17} $~&~$10^{-12} $~&~0.08~&~$0.90\%$~&~$-1.3\%$~\\
 		~~&~$ $~&~$ $~&~$10^{-18} $~&~$1.5\times10^{-13} $~&~0.03~&~$0.39\%$~&~$-0.58\%$~\\
 		\hline
 		\multirow{2}{*} ~NGC$104$~&~$2.3 \times 10^{3}$~&~$ 10^{12}$~&~$10^{-17} $~&~$6.5\times10^{-11} $~&~0.02~&~$0.21\% $~&~$-0.31\% $~\\
 		~~&~$ $~&~$ $~&~$10^{-18} $~&~$7.5\times10^{-12} $~&~0.01~&~$0.15\% $~&~$-0.23\% $~\\
 		\hline
 	\end{tabular}
 	\caption{Variations in the photon sphere and shadow radii, as estimated for a few candidate galactic black holes, for the constant mass density case. The DM parameters are only estimates based on current DM spike studies (for instance, \,\cite{2015Lacroix,2018Lacroix,2020Fortes}). }
 	\label{tab:mCPAsheff}
 \end{table}
 %%%%%%

%%%%%%%%%%%%%%%%%%%%%%%%%%%%%%%%%%%%%%%%%%%%%%%%%%%%%%%%%%%
\subsection{Radially in-falling dark matter plasma}
\label{sec:rinfsec}
We next turn to a model that is slightly more realistic, albeit still a very simplified one, that is amenable to analytic treatments. We consider now the scenario where the DM is radially in-falling into the black hole. Again, we assume that the DM plasma follows the total DM distribution.

For steady, radial accretion we have from the continuity equation
\begin{equation}
\partial_r\left(\sqrt{-g}\rho(r)u^r\right)=0\;.
\end{equation}
For the static, spherically symmetric case of the Schwarzschild geometry, we have from above, the equation
\begin{equation}
r^2\rho(r)\left(-\sqrt{\frac{2GM_{\text{\tiny{BH}}}}{r}}\right)=-\frac{\dot{M}_{ac}}{4\pi } \;.
\end{equation}
Here, $\dot{M}_{ac}$ is the accretion rate of the DM, into the central black hole.

This leads to a radial density distribution for the total DM
\begin{equation}
\rho(r)=\frac{\dot{M}_{ac}}{4\pi c \sqrt{R_{\text{\tiny{Sch.}}}\, r^3}}\;,
\end{equation}
assuming steady state accretion is occurring. 

For concreteness, let us assume that significant accretion is occurring only in a finite, spherically symmetric region around the black hole, again parametrised by a parameter $\kappa$. Then, we have for the DM plasma density distribution
\begin{equation}\label{eq:mCPAPRFFPD}
\tilde{\rho}(r)= \begin{cases}
\frac{f_{\text{\tiny mCP}}\dot{M}_{ac}}{4\pi c \sqrt{R_{\text{\tiny{Sch.}}}\, r^3}} \quad & R_{\text{\tiny{Sch.}}}<r<\kappa R_{\text{\tiny{Sch.}}} \;,\\
0 \quad & \kappa R_{\text{\tiny{Sch.}}}\le r \;. \\
\end{cases}
\end{equation}
The corresponding DM plasma frequency will have a functional dependence
\begin{equation}\label{eq:mCPAPRFFPF}
\tilde{\omega}^2_p(r)= \begin{cases}
\frac{\epsilon^2q_e^2 f_{\text{\tiny mCP}}\dot{M}_{ac}}{ 4\pi c \epsilon_om_{\tilde{e}} m_{\tilde{p}} \sqrt{R_{\text{\tiny{Sch.}}}\,r^3}} \quad & R_{\text{\tiny{Sch.}}}<r<\kappa R_{\text{\tiny{Sch.}}} \; , \\
0 \quad & \kappa R_{\text{\tiny{Sch.}}}\le r \; .\\
\end{cases}
\end{equation}
The mass distribution for this case will be given by
\begin{equation}\label{eq:mCPAPRFFMD}
M(r)= \begin{cases}
M_{\text{\tiny{BH}}} \quad & r\le R_{\text{\tiny{Sch.}}} \; ,\\ 
M_{\text{\tiny{BH}}}+\frac{2\dot{M}_{ac}}{3 c R_{\text{\tiny{Sch.}}}^{\frac{1}{2}}}\left(r^{\frac{3}{2}}-R_{\text{\tiny{Sch.}}}^{\frac{3}{2}}\right) \quad & R_{\text{\tiny{Sch.}}}<r<\kappa R_{\text{\tiny{Sch.}}} \;, \\ 
M_{\text{\tiny{BH}}}+\frac{2\dot{M}_{ac}R_{\text{\tiny{Sch.}}}}{3 c}\left(\kappa^{\frac{3}{2}}-1\right) \quad & \kappa R_{\text{\tiny{Sch.}}}\le r \; ,\\
\end{cases}
\end{equation}
and the correction to the $g_{tt}$ component of the metric due to this distribution is (also see Appendix A)
\begin{equation}\label{eq:mCPAPRFFf}
f(r)= \begin{cases}
1-\frac{R_{\text{\tiny{Sch.}}}}{r}-\frac{4G\dot{M}_{ac}}{3 c^3 R_{\text{\tiny{Sch.}}}^{\frac{1}{2}}r}\left(r^\frac{3}{2}-R_{\text{\tiny{Sch.}}}^\frac{3}{2}\right) \quad & R_{\text{\tiny{Sch.}}}<r<\kappa R_{\text{\tiny{Sch.}}} \;, \\
1-\frac{R_{\text{\tiny{Sch.}}}}{r}-\frac{4G\dot{M}_{ac}R_{\text{\tiny{Sch.}}}}{3 c^3r}\left(\kappa^{\frac{3}{2}}-1\right) \quad & \kappa R_{\text{\tiny{Sch.}}}\le r \;.\\
\end{cases}
\end{equation}

As before, let us define dimensionless parameters that will help quantify and demarcate the electromagnetic and gravitational effects due to the DM plasma component. In analogy to the constant density case, let us denote these distinct contributions by the following dimensionless parameters
\begin{eqnarray}\label{eq:mCPAPRFFp}
\chi_{d}&\equiv& \frac{\epsilon^2q_e^2f_{\text{\tiny mCP}} \dot{M}_{ac} }{4 \pi \epsilon_o c m_{\tilde{e}}m_{\tilde{p}} R_{\text{\tiny{Sch.}}}^2 E^2} \; ,\nonumber \\
\chi_{g}&\equiv& \frac{4 G \dot{M}_{ac}}{3 c^3} \; .
\end{eqnarray}
$\chi_d$ is again just sourced by the DM plasma, while $\chi_g$ depends on the total DM distribution.

With the above definitions, we have from Eqs.\,(\ref{eq:mCPAPRSH}), (\ref{eq:mCPAPRFFPF}) and (\ref{eq:mCPAPRFFf}) the expression for the modified shadow radius in the present case as
\begin{equation}\label{eq:mCPAPRFFrshe2}
r_{\text{\tiny{sh.}}}=\begin{cases}
&\sqrt{\left(1-\frac{R_{\text{\tiny{Sch.}}}}{r_{\text{\tiny{obs.}}}}-\chi_{g}\left(\left(\frac{r}{R_{\text{\tiny{Sch.}}}}\right)^\frac{1}{2}-\left(\frac{R_{\text{\tiny{Sch.}}}}{r}\right)\right)\right)\left(\frac{r_{\text{\tiny{ph.}}}^2}{1-\frac{R_{\text{\tiny{Sch.}}}}{r_{\text{\tiny{ph.}}}}}\right)}   ~~~;~~~~~ \kappa<3/2 \;,\\
\\
&\sqrt{\left(1-\frac{R_{\text{\tiny{Sch.}}}}{r_{\text{\tiny{obs.}}}}-\chi_{g}\frac{R_{\text{\tiny{Sch.}}}}{r_{\text{\tiny{obs.}}}}\left(\kappa^\frac{3}{2}-1\right)\right)\left(\frac{r_{\text{\tiny{ph.}}}^2}{1-\frac{R_{\text{\tiny{Sch.}}}}{r_{\text{\tiny{ph.}}}}-\chi_{g}\left(\left(\frac{r_{\text{\tiny{ph.}}}}{R_{\text{\tiny{Sch.}}}}\right)^\frac{1}{2}-\left(\frac{R_{\text{\tiny{Sch.}}}}{r_{\text{\tiny{ph.}}}}\right)\right)}-\chi_{d}r_{\text{\tiny{ph.}}}^2 \left(\frac{R_{\text{\tiny{Sch.}}}}{r_{\text{\tiny{ph.}}}}\right)^\frac{3}{2}\right)}  \\
&~~~~~~~~~~~~~~~~~~~~~~~~~~~~~~~~~~~~~~~~~~~~~~~~~~~~~~~~~~~~~~~~~~~~~~~~~~;~~~~~~~ \kappa\ge3/2 \;.\\
\end{cases}
\end{equation}

As in the case of constant mass density, note that the radius of the photon sphere is also modified. This new photon sphere radius will be now given implicitly by the criteria
\begin{equation}\label{eq:photoradiusifr}
\begin{split}
\frac{d}{dr}\left.\left[\frac{r^2}{1-\frac{R_{\text{\tiny{Sch.}}}}{r}-\chi_{g}\left(\left(\frac{r}{R_{\text{\tiny{Sch.}}}}\right)^\frac{1}{2}-\left(\frac{R_{\text{\tiny{Sch.}}}}{r}\right)\right)}\right]\right|_{r=r_{\text{\tiny{ph.}}}}=0 ~~~;~~~~~ \kappa<3/2 \;,\\
\frac{d}{dr}\left.\left[\frac{r^2}{1-\frac{R_{\text{\tiny{Sch.}}}}{r}-\chi_{g}\left(\left(\frac{r}{R_{\text{\tiny{Sch.}}}}\right)^\frac{1}{2}-\left(\frac{R_{\text{\tiny{Sch.}}}}{r}\right)\right)}-\chi_{d}r^2 \left(\frac{rs}{r}\right)^\frac{3}{2}\right]\right|_{r=r_{\text{\tiny{ph.}}}}=0 ~~~;~~~~~ \kappa \ge 3/2 \;.\\
\end{split}
\end{equation}\

\begin{figure}[H]
\center
\includegraphics[scale=0.6]{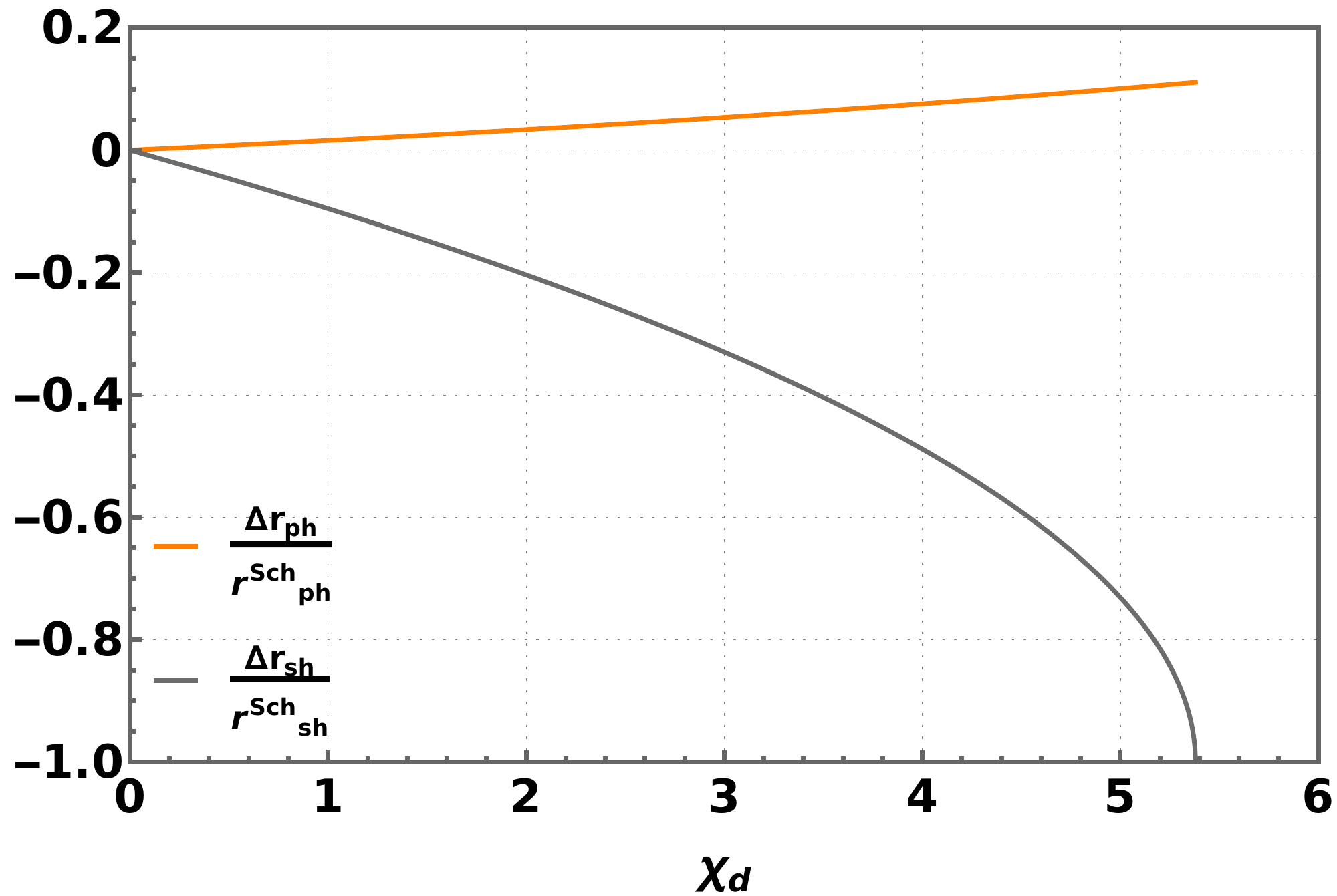}
\caption{The variations in the photon sphere and shadow radii with respect to the Schwarzschild blackhole values, in the presence of radially in-falling DM plasma. Again, here we have defined ${\Delta r_{\text{(\tiny{ph.,sh.})}}}/{r^{\text{\tiny{Sch.}}}_{\text{(\tiny{ph.,sh.})}}} =\left({ r_{\text{(\tiny{ph.,sh.})}}-r^{\text{\tiny{Sch.}}}_{\text{(\tiny{ph.,sh.})}}}\right)/{r^{\text{\tiny{Sch.}}}_{\text{(\tiny{ph.,sh.})}}}$. The photon sphere radius increases as $\chi_{d}$ increases. Upto $\sim$ 11\% change in the photon sphere radius can be observed in the presence of a DM plasma cloud. The shadow radius decreases  as $\chi_{d}$ increases. As $\chi_{d}$ approaches 5.379 the shadow radius approaches 0 and eventually vanishes. This happens because the impact parameter for photon trajectory approaches 0 as $\chi_{d}$ approaches 5.379. Note that beyond $\chi_d\simeq5.4$ there is no photonsphere.}
\label{fig:RIFchid}
\end{figure}

The above expressions may once more be be solved numerically. We display the corresponding variations in the photon sphere and shadow radii in Fig.\,\ref{fig:RIFchid}. For $\kappa<3/2$, any effect of the DM plasma cloud would appear only through the additional gravitational contribution of the total DM distribution. The photon sphere radius is found to again increase as a function of $\chi_d$, consistent with our observations in the constant DM density case. The black hole shadow radius in contrast is found to again decrease with increasing $\chi_d$. This may be traced to the refractive contribution of the in-falling DM plasma that has a defocusing effect on the outgoing light rays. As we derive below, for $\chi_d\rightarrow5.4$, the DM plasma becomes opaque to the electromagnetic wave grazing the photon sphere; note again from Eq.\,(\ref{eq:mCPAPRFFp}) that $\chi_d$ implicitly depends on $\lambda_{\text{\tiny{EM}}}$. On the other hand, the photon sphere itself does not exist beyond $\chi_d\simeq5.4$ in the present case, as there are no real solutions to $r_{\text{\tiny{ph.}}}$.

%%%%%
%%%%%
To get an analytic understanding of the result and the relative influence of the various contributions, we may once again consider a regime where $\chi_{g,d}\ll 1$. Solving Eq.\,(\ref{eq:photoradiusifr}) perturbatively in the radially in-falling DM plasma case then gives for the modified photon sphere radius
\begin{equation}\label{eq:mCPAPRFFrp}
\frac{r_{\text{\tiny{ph.}}}~}{R_{\text{\tiny{Sch.}}}}\simeq \frac{3}{2}+\frac{\chi_{d} }{18 \sqrt{6}}+\frac{7 \chi_{d} ^2}{5832}+\frac{85 \chi_{d} ^3}{629856 \sqrt{6}}+\frac{3}{16}\left(-8+3 \sqrt{6}\right) \chi_{g} +\frac{1}{432} \left(-27+2 \sqrt{6}\right) \chi_{d}  \chi_{g} + O(\chi_{g,d}^3)\; .\\
	\end{equation}
Using the above expression, one then obtains the corresponding shadow radius from Eq.\,(\ref{eq:mCPAPRFFrshe2}) in the $\chi_{g,d}\ll 1$ regime as 
\begin{eqnarray}\label{eq:mCPAPRFFrsh}
\frac{r_{\text{\tiny{sh.}}}}{R_{\text{\tiny{Sch.}}}} &\simeq& \frac{3 \sqrt{3}}{2}\left(1-\frac{R_{\text{\tiny{Sch.}}}}{r_{\text{\tiny{obs.}}}}\right)^{\frac{1}{2}}-\frac{\chi_{d} }{3 \sqrt{2}}\left(1-\frac{R_{\text{\tiny{Sch.}}}}{r_{\text{\tiny{obs.}}}}\right)^{\frac{1}{2}}\nonumber \\
&-&\frac{3\sqrt{3}\chi_{g}}{8}\left(3\sqrt{6}\left(1-\frac{r_{\text{\tiny{obs.}}}}{R_{\text{\tiny{Sch.}}}}\right)+4\frac{r_{\text{\tiny{obs.}}}}{R_{\text{\tiny{Sch.}}}}-6+2\kappa^{\frac{3}{2}}\right)\left(\frac{R_{\text{\tiny{Sch.}}}^2}{r_{\text{\tiny{obs.}}}(r_{\text{\tiny{obs.}}}-R_{\text{\tiny{Sch.}}})}\right)^{\frac{1}{2}}\\
&+&\mathcal{O}(\chi_{g,d}^2) \nonumber \; .
\end{eqnarray}
 
 As in the constant mass density case let us quantify the typical magnitudes for the dimensionless parameters $\chi_g$ and $\chi_d$ for representative black hole and related astrophysical parameters. From Eq.\,(\ref{eq:mCPAPRFFp}), the typical values that these dimensionless parameters may take can be quantified as
\begin{eqnarray}\label{eq:mCPAPRFFp1}
\chi_{d}&=&4.93\times10^{26} \times\frac{\epsilon^2}{\eta} \times \left(\frac{f_{\text{\tiny mCP}}\dot{M}_{ac}}{\solarmass \mathrm{yr}^{-1}  }\right) \times \left(\frac{\mathrm{eV}/c^2}{ m_{\tilde{p}}}\right)^2 \times\left(\frac{\solarmass}{M_{\text{\tiny{BH}}} }\right)^2 \times \left(\frac{\lambda_{\text{\tiny{EM}}}}{ \mathrm{mm} }\right)^2 \; , \nonumber \\
\chi_{g}&=& 2.08\times10^{-10}\times\left(\frac{\dot{M}_{ac}}{\solarmass \mathrm{yr}^{-1}  }\right) \; .
\end{eqnarray}

Similar to the constant mass density case, as may be expected, the gravitational effect of the DM plasma is again largely insignificant, for reasonable mCP and astrophysical parameters. For instance, taking $f_{\text{\tiny mCP}} \dot{M}_{ac}=0.1 \solarmass/\mathrm{Year}$, with $f_{\text{\tiny mCP}}=0.1\%$, which is comparable to typical baryonic accretion rates near black holes in many cases, one obtains $\chi_{g}\simeq2.08\times10^{-11}$. This value for $\chi_g$ is insignificant and its concomitant effects in Eqs.\,(\ref{eq:mCPAPRFFrshe2}) and (\ref{eq:photoradiusifr}) will be negligible as well. The typical $\chi_d$ values may again be significant. For $\eta=1$, $\lambda_{\text{\tiny{EM}}}=1.3\,\mathrm{mm}$, $\epsilon=10^{-17}$ and $m_{\text{\tiny mCP}}=10^{-13}$ eV, we get $\chi_d\sim\mathcal{O}(0.01)$, for instance. For Hydrogen plasma, the corresponding $\chi_{d}$ is very small, again around  $10^{-9}$. Large mCP masses again only lead to small effects. Unlike in the constant mass density case, we are unable to find a closed form analytic expression for the photon sphere and shadow radii by taking $\chi_{g}=0$.

Let us now address the attenuation of electromagnetic radiation in the present case. For  viable transmission of an electromagnetic wave at a radial distance `r', we require the necessary condition $\tilde{\omega}^2_p(r)<\omega^2(r)$, which in the radially in-falling scenario translates to
\begin{equation}
\chi_{d}(r) \left(\frac{R_{\text{\tiny{Sch.}}}}{r}\right)^\frac{3}{2}<\frac{1}{f(r)} \; .
\end{equation}
As $\chi_{g}$ is mostly negligible, for realistic parameters, we may approximate the above condition to be a requirement on just the $\chi_d(r)$. This gives
\begin{equation}\label{eq:inremtrans}
\chi_{d}(r)<\frac{\left(\frac{r}{R_{\text{\tiny{Sch.}}}}\right)^\frac{3}{2}}{1-\frac{R_{\text{\tiny{Sch.}}}}{r}} \; ,
\end{equation}
as the necessary criterion to have non attenuated transmission of electromagnetic waves at a point. Since $\chi_d$ implicitly has a factor of $\lambda_{\text{\tiny{EM}}}$ in its definition, the necessary condition on the electromagnetic radiation may also equivalently be written as
\begin{equation}\label{eq:inremtranswave}
\lambda_{\text{\tiny{EM}}} < \left(\frac{16 \pi^3 c^3 \epsilon_o m_{\tilde{e}}m_{\tilde{p}} R_{\text{\tiny{Sch.}}}^{\frac{1}{2}}r^{\frac{3}{2}} }{\epsilon^2q_e^2f_{\text{\tiny mCP}}\dot{M}_{ac}\left(1-\frac{R_{\text{\tiny{Sch.}}}}{r}\right)}\right)^{\frac{1}{2} }\;.
\end{equation}

%%%%%
\begin{figure}[H]
	\center
	\subfloat[]
	{\includegraphics[scale=0.175]{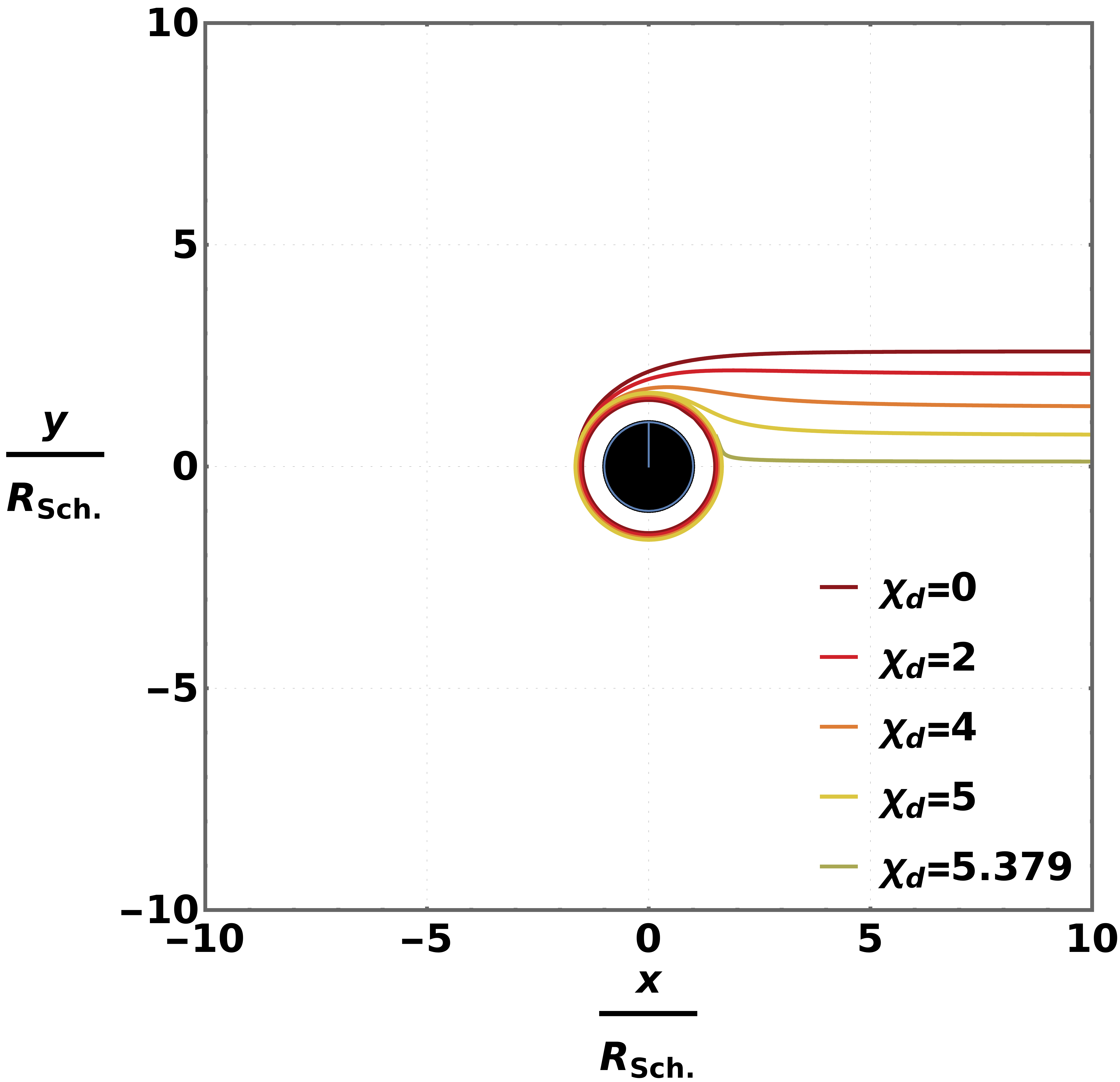}}
	\subfloat[]
	{\includegraphics[scale=0.192]{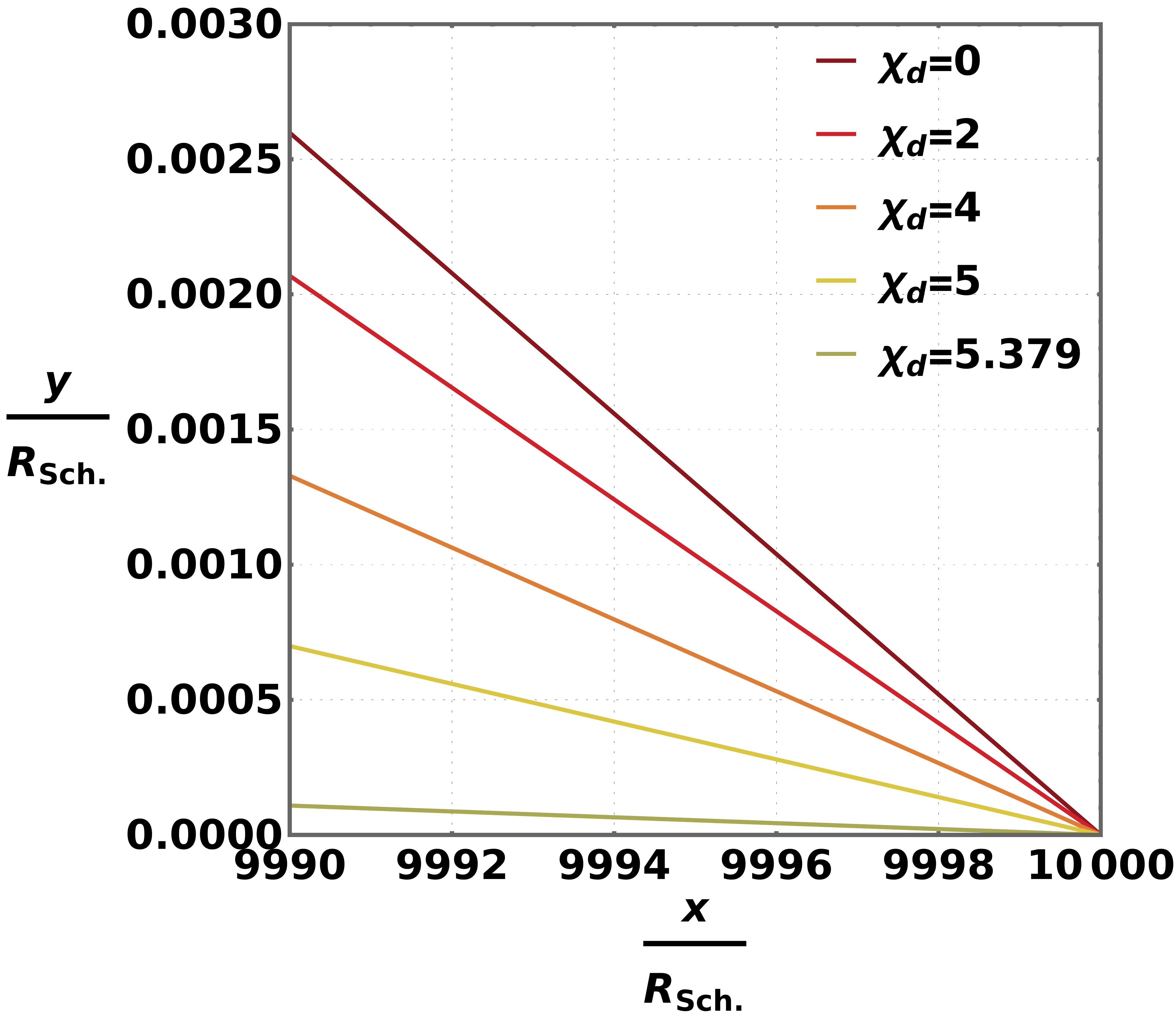}}\\
	\subfloat[]
	{\includegraphics[scale=0.275]{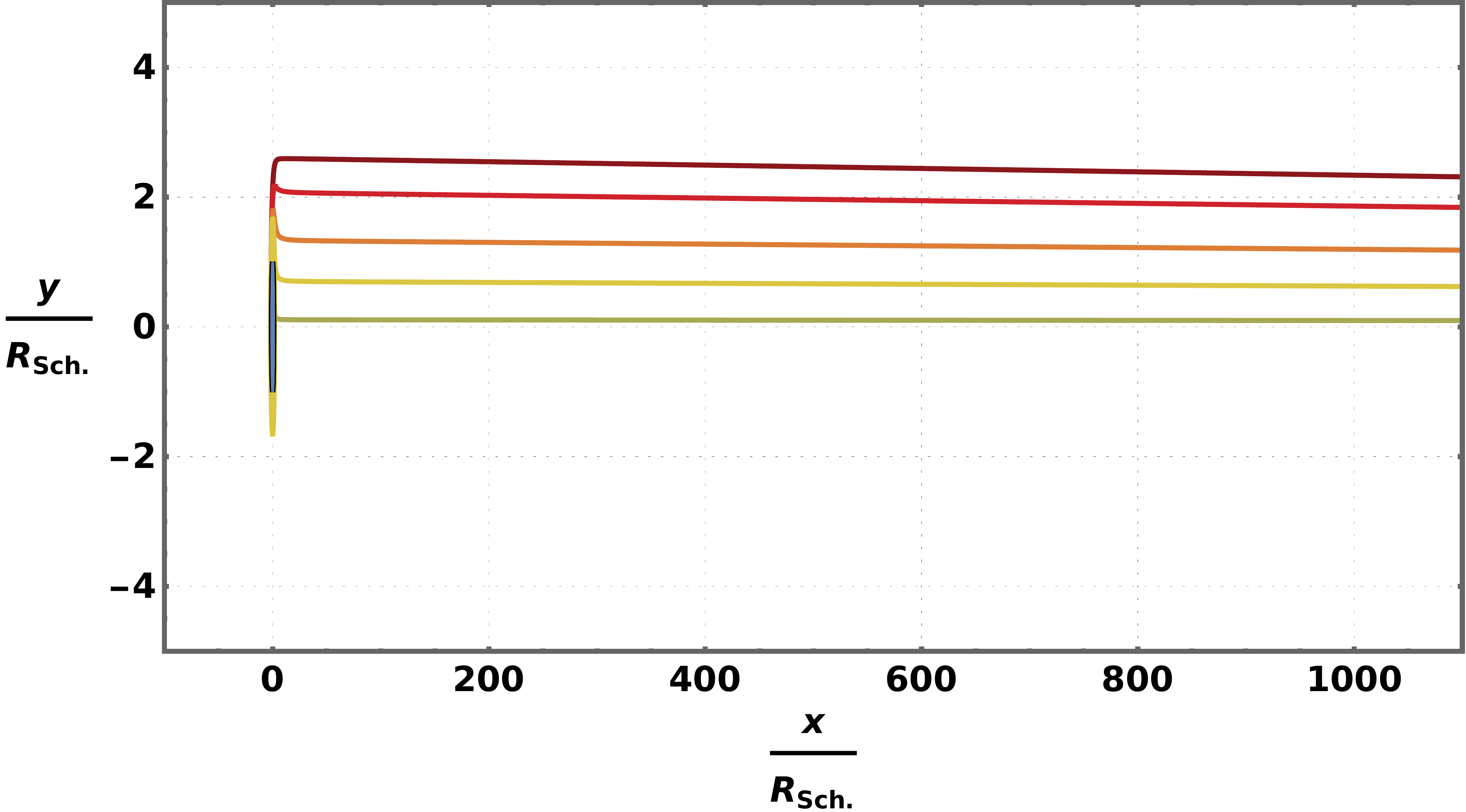}
		\includegraphics[scale=0.275]{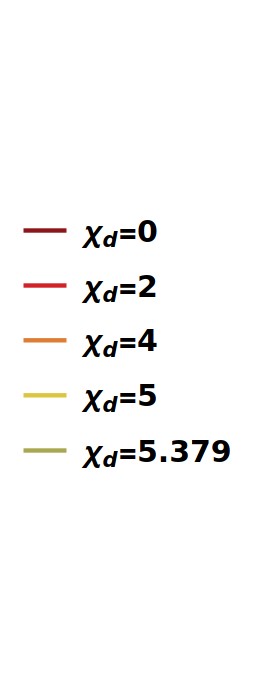}	}
	\caption{The outgoing light ray trajectories are shown in the equatorial plane of the blackhole, for the case of in-falling DM plasma. We have taken $f_{\text{\tiny mCP}} \dot{M}_{ac}=0.1 \solarmass/\mathrm{Year}$, $M_{\text{\tiny BH}}=6.5\times10^9 \solarmass$, $\lambda=1.3\,\mathrm{mm}$ and $\kappa=10^3$. $(x/R_{\text{\tiny{Sch.}}},y/R_{\text{\tiny{Sch.}}})=(0,0)$ is the center of the blackhole and the observer is located at $(x/R_{\text{\tiny{Sch.}}},y/R_{\text{\tiny{Sch.}}})=(10^4,0)$. The outgoing null rays are shown for various values of $\chi_d$. We have again focused on three distinct regions to illustrate the nature of the light ray trajectories. In subfigure (a) we can just make out the increase in photon sphere radius with increasing values of $\chi_d$. From subfigure (b), we clearly see that from the perspective of the observer's cone of avoidance, as $\chi_d\rightarrow 5.379$, the shadow radius vanishes.}
	\label{fig:RIFlrt}
\end{figure}
%%%%%	
For example, considering $f_{\text{\tiny mCP}}\, \dot{M}_{ac} =0.1 \solarmass/\mathrm{Year}$, $M_{\text{\tiny BH}}=6.5\times10^9 \solarmass$, $m_{\tilde{e}}=m_{\tilde{p}}$, $r=\frac{5}{3} R_{\text{\tiny{Sch.}}}$ and taking $\epsilon=10^{-17}$, and $m_{\tilde{p}}= 10^{-15}$ eV, we obtain $\lambda_{\text{\tiny{EM}}}<0.21$ mm. Similarly, for $\epsilon=10^{-17}$ and $m_{\tilde{p}}=10^{-13}$ eV, we obtain $\lambda_{\text{\tiny{EM}}}<21.
46$ mm.
Since the minimum value of the right-hand side of Eq.\,(\ref{eq:inremtrans}) occurs at $r=\frac{5}{3} R_{\text{\tiny{Sch.}}}$, we may place a necessary condition for the viable transmission of photon sphere grazing null trajectories as
\begin{equation}\label{eq:inrviabletranschid}
 \chi_{d}<5.379\;.
\end{equation}
Beyond this value, the photon sphere also does not exist. This is because, beyond this value of $\chi_d$, there are no positive real solutions to Eq.\,(\ref{eq:mCPAPPSR}).

Let us now discuss few of the theoretical features in some detail. As already observed, $\chi_{g}\ll1$ and the gravitational effect due to the radially in-falling DM has no appreciable influence on the photon sphere and shadow radii. For small value of $\chi_{d}$, Eqs.\,(\ref{eq:mCPAPCDrp}) and (\ref{eq:mCPAPCDrsh}) imply that the variation in the photon sphere and shadow radii due to the DM plasma should be negligible. In these regimes, as should be expected, one also recovers from the exact numerical computations, the Schwarzschild values
 \begin{equation}
 \left.\frac{r_{\text{\tiny{ph.}}}~}{R_{\text{\tiny{Sch.}}}}\right|_{\chi_{d}\ll1}\simeq\frac{3}{2},~\hquad\left.\frac{r_{\text{\tiny{sh.}}}~}{R_{\text{\tiny{Sch.}}}}\right|_{\chi_{d}\ll1}\simeq\frac{3\sqrt{3}}{2}\,.
 \end{equation}
 
  As in the constant mass density case, when we vary $\chi_d$, the DM plasma causes a change in the effective potential of Eq.\,(\ref{eq:mCPAPV}). The radial position of the effective potential's maxima gets shifted to higher values. This as before translates to an increase in the photon sphere radius as $\chi_{d}$ increases. This is the reasoning for the observed increase of the photon sphere radius in Fig.\,\ref{fig:RIFchid}, for larger $\chi_d$ values. 
 
 However, compared to the constant mass density case, the shift in the photon sphere radius is relatively small in the in-falling DM plasma case, for viable mCP and realistic astrophysical parameters. As we noted in Eq.\,(\ref{eq:inrviabletranschid}), beyond $\chi_d=5.379$, the electromagnetic transmission through the DM plasma is completely attenuated for the photon trajectories skimming the photon sphere . As $\chi_d \rightarrow 5.379$, the photon sphere radius approaches a fixed value of
 \begin{equation}
 \left.{r_{\text{\tiny{ph.}}}}\right|_{\chi_{d}\to5.379}\to\frac{5}{3}~{R_{\text{\tiny{Sch.}}}}\, .
 \end{equation} 
 As mentioned earlier, beyond $\chi_d\simeq 5.379$ the photon sphere also does not exist, as there are no positive real solutions to Eq.\,(\ref{eq:mCPAPPSR}), with $-\tilde{V}_{eff}$ in Eq.\,(\ref{eq:mCPAPV}) positive.
 
 Similar to the constant mass density case, in the radially in-falling case there is a net decrease in the shadow radius as $\chi_d$ increases. The general trend in the variation of the shadow radius is therefore consistent with the observation in the constant mass density case. This reinforces the argument that the general features are robust and not mere artefacts dependent on our astrophysical modelling of DM.
 
 But, in the in-falling case there is a remarkable observation that is distinct from the constant mass density case. From Eq.\,(\ref{eq:mCPAPCDsha}) it is observed, that the shadow radius decreases from its initial value $3\sqrt{3}/{2}$R$_{\text{\tiny{Sch.}}}$ and approaches 0  as $\chi_{d}$ approaches 5.379. This means that the black hole shadow \textit{disappears} completely,
 \begin{equation}
 \left.{r_{\text{\tiny{sh.}}}}\right|_{\chi_{d}\to5.379}\to0\, .
 \end{equation}
 
 This interesting feature is also verified from analysing the light ray trajectories in the spacetime. Fig.\,\ref{fig:RIFlrt} shows the outgoing light ray trajectories (i.e. towards the observer) in the radially in-falling DM plasma case. For clarity, we display these trajectories separately in three regions---vicinity of the blackhole, near the observer and the neighbourhood of the black hole. The $\chi_{d}\rightarrow 5.379$ case and the complete vanishing of the shadow is seen clearly from Fig.\,\ref{fig:RIFlrt} . The impact parameter for the light ray trajectories is vanishing as $\chi_{d}\rightarrow 5.379$. It may appear astonishing at first that despite the event horizon located at ${R_{\text{\tiny{Sch.}}}}$, which should be seen as a dark region, from the perspective of the distant observer one can still see some photons apparently emanating from the central regions. This is of course just due to the non-geodesic deviation of the light ray trajectories due to refraction, as the in-falling DM plasma acts akin to a lens.  
 
 %%%%%%   
\begin{table}[H]
	\center
	\renewcommand{\arraystretch}{1.25}
	\begin{tabular}{|c|c|c|c|c|c|c| } 
		\hline 
		~$\epsilon$~&~$m_{\text{\tiny mCP.}}~(eV)$~&~$\chi_{d}$~& $r_{\text{\tiny{ph.}}}~ (R_{\text{\tiny{Sch.}}})$~& ~$\frac{\Delta r_{\text{\tiny{ph.}}}}{r^{\text{\tiny{Sch.}}}_{\text{\tiny{ph.}}}} $~& $r_{\text{\tiny{sh.}}}~ (R_{\text{\tiny{Sch.}}})$~& ~$\frac{\Delta r_{\text{\tiny{sh.}}}}{r^{\text{\tiny{Sch.}}}_{\text{\tiny{sh.}}}}$~\\
		\hline
		\multirow{1}{*}{$10^{-14}$}&~$10^{-10}$~&~0.02~&~$1.500$~&~$\sim 0\%$~&~$2.593$~&~$-0.18\%$~\\
		\hline
		\multirow{2}{*}{$10^{-15}$}&~$10^{-11}$~&~0.02~&~$1.500$~&~$\sim 0\%$~&~$2.593$~&~$-0.18\%$~\\
		&~$5\times10^{-12}$~&~0.08~&~$1.502$~&~$ 0.12\%$~&~$2.579$~&~$ -0.72\%$~\\
		\hline
		\multirow{3}{*}{$10^{-16}$}&~$10^{-12}$~&~0.02~&~$1.500$~&~$\sim 0\%$~&~$2.593$~&~$-0.18\%$~\\
		&~$5\times10^{-13}$~&~0.08~&~$1.502$~&~$ 0.12\%$~&~$2.579$~&~$ -0.72\%$~\\
		&~$1.5\times10^{-13}$~&~0.88~&~$1.521$~&~$1.4\%$~&~$2.382$~&~$ -8.3\%$~\\
		\hline
		\multirow{3}{*}{$10^{-17}$}&~$10^{-13}$~&~0.02~&~$1.500$~&~$\sim 0\%$~&~$2.593$~&~$-0.18\%$~\\
		&~$1.5\times10^{-14}$~&~0.88~&~$1.521$~&~$1.4\%$~&~$2.382$~&~$-8.3\%$~\\
		&~$10^{-14}$~&~1.97~&~$1.550$~&~$ 3.3\%$~&~$2.077$~&~$-20\%$~\\
		\hline
		\multirow{5}{*}{$2\times10^{-18}$}&~$10^{-14}$~&~0.08~&~$1.502$~&~$0.12\%$~&~$2.579$~&~$-0.72\%$~\\
		&~$5\times10^{-15}$~&~0.32~&~$1.507$~&~$0.49\%$~&~$2.523$~&~$-2.9\%$~\\
		&~$1.5\times10^{-15}$~&~3.51~&~$1.597$~&~$ 6.5\%$~&~$1.545$~&~$- 41\%$~\\
		&~$1.25\times10^{-15}$~&~5.05~&~$1.653$~&~$10\%$~&~$0.649$~&~$-75\%$~\\
		&~$1\times10^{-15}$~&~7.88~&(No photon sphere)&$-$&~(No shadow)~&~$-$~\\
		\hline
	\end{tabular}
\caption{Variation in the photon sphere and shadow radii for few representative points in the physically viable $\epsilon-m_{\text{\tiny mCP}}$ parameter space, for the case of radially in-falling DM plasma. We have taken $f_{\text{\tiny mCP}} \dot{M}_{ac}=0.1 \solarmass/\mathrm{Year}$, $M_{\text{\tiny BH}}=6.5\times10^9 \solarmass$, $\kappa=10^3$, $\eta=1$ and $\lambda=1.3\,\mathrm{mm}$. In the table, we have defined ${\Delta r_{\text{(\tiny{ph.,sh.})}}}/{r^{\text{\tiny{Sch.}}}_{\text{(\tiny{ph.,sh.})}}} =\left({ r_{\text{(\tiny{ph.,sh.})}}-r^{\text{\tiny{Sch.}}}_{\text{(\tiny{ph.,sh.})}}}\right)/{r^{\text{\tiny{Sch.}}}_{\text{(\tiny{ph.,sh.})}}}$. Generally, for higher $\epsilon$ and lower $m_{\text{\tiny mCP.}}$, the change in photon sphere and shadow radii are larger, as expected from Eq.\,(\ref{eq:mCPAPRFFp1}). For most parts of the $\epsilon-m$ parameter space, the change in the photon sphere and shadow radii are insignificant. However, there are regions where the change in these radii are significant and potentially observable. Note that for $\epsilon\sim 10^{-18}$, in the neighbourhood of $m_{\text{\tiny{mCP}}}=5\times10^{-15}$, there is a steep increase in the photon sphere and shadow radii. This is a consequence of the refractive effect of the DM plasma becoming significant (i.e. $\chi_{d}\sim\mathcal{O}(1)$) in the vicinity of these parameter space regions. In these regions, the DM plasma term in the effective potential of Eq.\,(\ref{eq:mCPAPV}) starts to become important. }
\label{tab:mCPAPRIFem}
\end{table}
%%%%%%
%%%%%%   
 \begin{table}[H]
 	\center
 	\renewcommand{\arraystretch}{1.25}
 	\begin{tabular}{|c|c|c|c|c|c|c|c| } 
 		\hline 
 		Blackhole&Mass $(\solarmass)$ & $f_{\text{\tiny  mCP}} \dot{M}_{ac}\,\left(\mathrm{\solarmass/ Year}\right) $&$\epsilon$&$m_{\text{\tiny mCP}}\,\mathrm{(eV)}$&$\chi_{d}$&$\frac{\Delta r_{\text{\tiny{ph.}}}}{r^{\text{\tiny{Sch.}}}_{\text{\tiny{ph.}}}} $&$\frac{\Delta r_{\text{\tiny{sh.}}}}{r^{\text{\tiny{Sch.}}}_{\text{\tiny{sh.}}}}$\\ 
 		\hline
 		\multirow{2}{*} ~M$87$*~&~$6.5 \times 10^{9}$~&~$1 $~&~$10^{-17} $~&~$2.5\times10^{-14} $~&~3.16~&~$ 5.7\%$~&~$ -35\%$~\\
 		~&~$ $~&~$ $~
 		&~$10^{-18} $~&~$2\times10^{-15} $~&~4.93~&~$ 9.9\%$~&~$ -71\%$~\\
 		\hline
 		\multirow{2}{*} ~Sgr A*~&~$4.3 \times 10^{6}$~&~$10^{-3} $~&~$10^{-17} $~&~$10^{-12} $~&~4.51~&~$8.8\%$~&~$-59\% $~\\
 		~&~$ $~&~$  $~&~$10^{-18} $~&~$1.5\times10^{-13} $~&~2.00~&~$3.4\%$~&~$-20\% $~\\
 		\hline
 		\multirow{2}{*} ~NGC$104$~&~$2.3 \times 10^{3}$~&~$ 10^{-6}$~&~$10^{-17} $~&~$6.5\times10^{-11} $~&~3.73~&~$7\% $~&~$-44\% $~\\
 		~ ~&~$ $~&~$  $~&~$10^{-18} $~&~$7.5\times10^{-12} $~&~2.8~&~$5.0\% $~&~$-30\% $~\\
 		\hline
 	\end{tabular}
 	\caption{Variations in the photon sphere and shadow radii for a few candidate galactic black holes, for the case of radially in-falling DM plasma. The DM parameters are again only estimates based on existing DM spike literature\,\cite{2015Lacroix,2018Lacroix,2020Fortes}.}
 	\label{tab:mCPAsheffir}
 \end{table}
 %%%%%%%%%%%
 We list a few representative points in the $\epsilon-m_{\text{\tiny mCP}}$ parameter space in Table \ref{tab:mCPAPRIFem}. We have taken, as before,  $f_{\text{\tiny mCP}}=0.1\%$ consistent with Eq.\,(\ref{eq:bcc}) and  assumed then that $f_{\text{\tiny mCP}}\dot{M}_{ac}=0.1\, \solarmass/\mathrm{Year}$. For the latter value we are assuming that the DM accretion is at least comparable to typical baryonic plasma accretions that are observed. As before, we take $M_{\text{\tiny BH}}=6.5\times10^9 \solarmass$, $\eta=m_{\tilde{e}}/m_{\tilde{p}}=1$ and $\lambda=1.3\, \mathrm{mm}$\,\cite{Akiyama:2019cqa}. In the physically viable regions of the mCP parameter space, as summarised in Sec.\,\ref{sec:mcps}, we see that while in most regions the effects are small, there are nevertheless regions where the variations are significant and potentially observable. For example, at  $\epsilon=10^{-17}$ and $m_{\text{\tiny mCP}}= 10^{-14}\,\mathrm{eV}$ relative change in photon sphere radius is around $3.3\%$ and that in the shadow radius is $20\%$.
 
 %%%%%%%%%%%
 \begin{figure}[H]
 	\center
 	\includegraphics[scale=0.475]{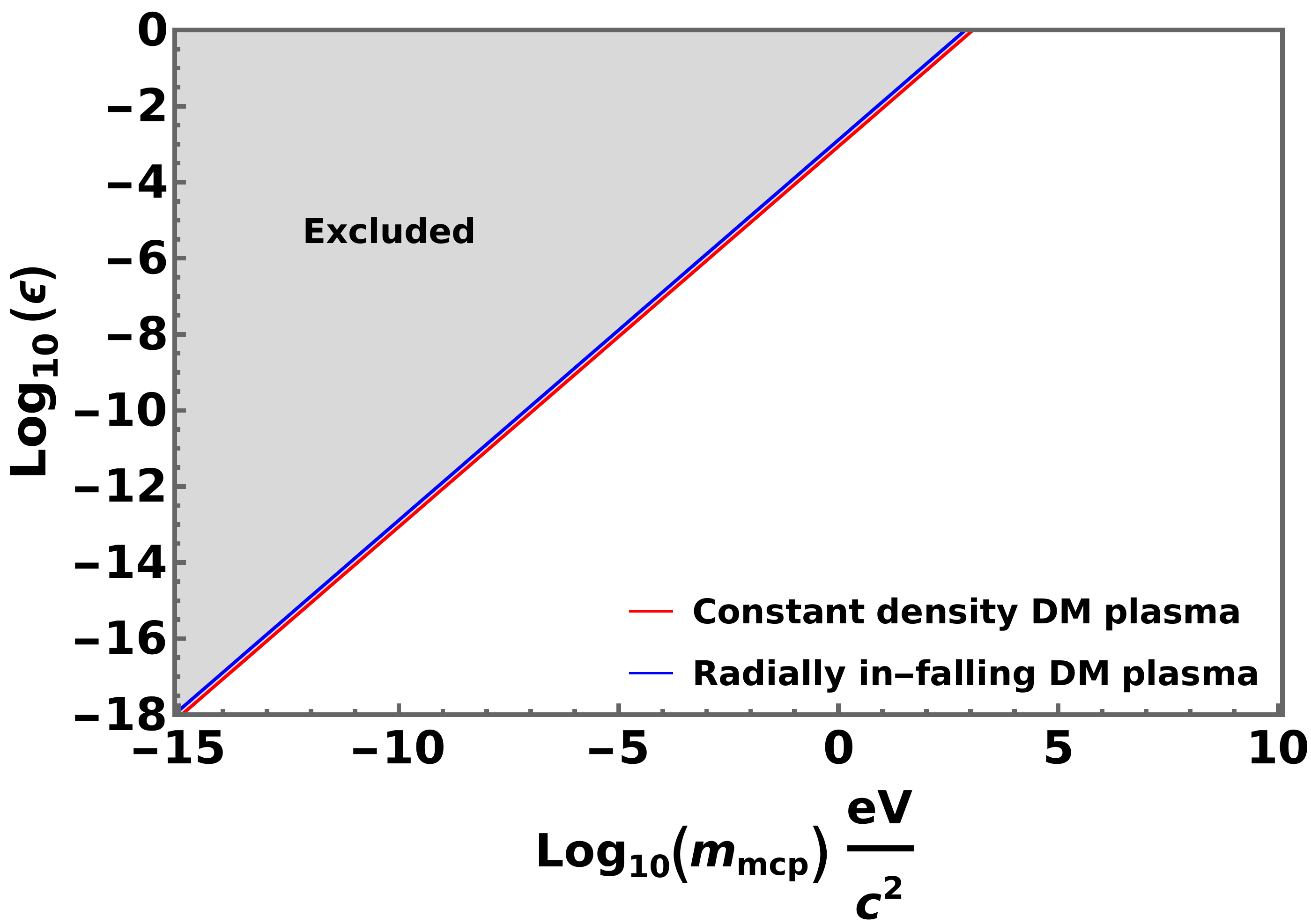}
 	\caption{Regions of the mCP parameter space that are excluded if one requires that $\Delta r_{\text{\tiny{sh.}}}/r^{\text{\tiny{Sch.}}}_{\text{\tiny{sh.}}}<20\%$ and that the DM plasma remains transparent to $\lambda_{\text{\tiny{EM}}}=1.3\,\mathrm{mm}$ electromagnetic waves, with no opacity features in astronomical observations. The exclusion lines for both the constant mass density as well as the radially in-falling cases are shown. The isoclines are almost coincident, suggesting that that the result is largely independent of astrophysical modelling. We are assuming the same astrophysical parameters as those in Tables\,\ref{tab:mCPAPCDem} and\,\ref{tab:mCPAPRIFem}. If future detailed observations\,\cite{Akiyama:2019cqa,ngEHT,WebbSgrA} do not see any substantial deviations, from normal expectations, we see that interesting regions of the mCP parameter space may be disfavored. }
 	\label{fig:cdm_rif_constraint}
 \end{figure}
 %%%%%
It is observed that the shadow radius decreases as $\epsilon$ increases or $m_{\text{\tiny mCP}}$ decreases, but the photon sphere radius follows an opposite trend i.e. it increases with increasing $\epsilon$ or decreasing $m_{\text{\tiny mCP}}$. The maximum change in the shadow radius observed, in viable regions of the mCP parameter space, may be as high as $100\%$. For some region of the allowed $\epsilon-m_{\text{\tiny mCP}}$ parameter space, the photon sphere grazing null ray gets completely attenuated by the DM plasma as $\chi_d \rightarrow 5.379$.

Again, in Table \ref{tab:mCPAsheffir} we show estimates for a few galactic black holes. We are again neglecting the spins of the black holes. We note that compared to the constant mass density case, with the parameters assumed, one obtains a larger variation. Here, as in the constant mass density case, the caveat is that the DM plasma distribution and accretion parameters are only crude estimates\,\cite{2015Lacroix,2018Lacroix,2020Fortes}. Nevertheless, the possibility that in many regions of the mCP parameter space we may have potentially interesting variations in the near-horizon features and shadow radius persists.

Even more precise observations in future, of black hole shadows and their neighbourhoods\,\cite{Akiyama:2019cqa,ngEHT,WebbSgrA}, may place bounds on how much the shadow radii may deviate from normal expectations in the Schwarzschild or Kerr cases. Fig.\,\ref{fig:cdm_rif_constraint} shows the excluded regions in the mCP parameter space if one requires that the shadow radius in the spherically symmetric case does not deviate by more than $20\%$ from normal expectations, and that there are no DM plasma induced opacity features in the astronomical observations. The exclusion lines from the constant mass density and radially in-falling scenarios are both shown. Parameter values are being assumed consistent with Tables\,\ref{tab:mCPAPCDem} and\,\ref{tab:mCPAPRIFem}. We see that the exclusion isoclines are in close proximity and may in principle rule out interesting mCP regions if precise observations of the near-horizon regions become possible in the near-future.

 %%%%%%%%%%%%%%%%%%%%%%%%%%%%%%%%%%%%%%%

\section{Summary and conclusions}
\label{sec:summary}
In this work, we studied the potential effects of a component of dark matter in the form of millicharged constituents, a putative dark matter plasma component, surrounding a spherically symmetric black hole. The extreme gravitational potential well near the black hole could lead to local dark matter over-densities\,\cite{Gondolo_1999,Lacroix2015,nampalliwar2021modelling,Feng:2021qkj,Kim:2021yyo} and the dark matter plasma component may then have non-trivial effects on the near-horizon characteristics and black shadows. 

We first considered the physically viable parameter space for millicharged particles and co-opted some of the methods from previous works\,\cite{2004,Feng:2009mn,PhysRevD.85.101302} to estimate what the fraction of a dark matter plasma could be in the parameter space of interest to us. We then proceeded to theoretically analyse in detail the millicharged particle furnished dark matter plasma, in the two-fluid approximation, assuming it is cold, pressure-less and non-magnetised. In the same context, we considered two semi-realistic astrophysical scenarios, corresponding to a constant mass density distribution for the dark matter plasma and a scenario where the dark matter plasma was assumed to be radially in-falling into the black hole. We derived the theoretical expressions quantifying the dark matter plasma effects in these specific cases, for various regions of the viable parameter space, and then studied comprehensively the unique variations that the photon sphere and black hole shadow radii could be subjected to. 

We find that while in most regions of the viable millicharged particle parameter space the effects are small, there are interesting regions where the effects could be very significant. Apart from the theoretical results in Secs.\,\ref{subsec:imps}, \ref{sec:cmdsec} and \ref{sec:rinfsec}, some of our main results are contained in Figs.\,\ref{fig:CMDchid}-\ref{fig:cdm_rif_constraint} and Tables\,\ref{tab:mCPAPCDem}-\ref{tab:mCPAsheffir}. 

Many of the observed features and variations are markedly distinct, from Hydrogen plasma effects that have been previously explored in the literature\,\cite{PhysRevD.92.104031,Chowdhuri:2020ipb,li2021gravitational,Badia:2021kpk,perlick2000ray,Atamurotov:2021hoq}, and electromagnetically neutral perfect fluid dark matter studies with phantom field-like equations of state considered earlier\,\cite{Xu:2017bpz,Haroon:2018ryd,Hou:2018avu,Jusufi:2019ltj,Shaymatov:2020bso,Saurabh:2020zqg,Badia:2020pnh,Rayimbaev:2021kjs,Atamurotov:2021hck}. We find that the the photon sphere radius increases and the shadow radius decreases as the dark matter plasma effect becomes more pronounced, as a function of the millicharged particle mass and charge. In the constant mass density case, it is found that the relative increase in the shadow radius can be as large as $\sim 25\%$, and for radially in-falling dark matter plasma even as large as $\sim 100\%$; all within the physically viable millicharged particle parameter space. In both cases, there are also critical values for the dark matter plasma dispersion parameter---beyond which there is no photon sphere and where the the dark matter plasma becomes opaque to electromagnetic radiation of a particular wavelength. In the radially in-falling dark matter plasma case, for certain values of the millicharged particle mass and charge, the shadow radius vanishes, from the perspective of the distant observer. 

A study of the corresponding light ray trajectories in all cases confirm the theoretical results. We also estimated putative effects of a dark matter plasma component on some of the currently observed galactic black holes, neglecting their spin, and find that in some scenarios the effects may be observable in future. Postulating further on future precise observations of black hole shadows and neighbourhoods\,\cite{Akiyama:2019cqa,ngEHT,WebbSgrA} and their potential ability to constrain possible deviations from normal expectations in the Schwarzschild case, we also speculated on possible exclusion limits that may be placed in the millicharged particle parameter space.

The conservative results we obtain, with reasonable assumptions for the millicharged particle sourced dark matter plasma and astrophysical scenarios, suggest that if such species exist and form a component of dark matter, they may be amenable to detection by their indirect influence on black hole near-horizon characteristics. Alternatively, an absence of such deviations in future exquisite observations can potentially place interesting limits on the millicharged particle parameter space.

%%%%%%%%%%%%%%%%%%%%%%%%%%%%%%%%%%%%%%%%%%%%%%%%%%%%%%%%%%%%%

\acknowledgments
We are grateful to R. Bhalerao and P. Subramanian for useful discussions. LB would also like to thank S. Khan for discussions. LB acknowledges support from a Junior Research Fellowship, granted by the Human Resource Development Group, Council of Scientific and Industrial Research, Government of India. AT would like to acknowledge support from an Early Career Research award, from the Department of Science and Technology, Government of India.

%%%%%%%%%%%%%%%%%%%%%%%%%%%%%%%%%%%%%%%%%%%%%%%%%%%%%%%%%%%%%
\appendix
%%%%%%%%%%%%%%%%%%%%%%%%%%%%%%%%%%%%%%%%%%%%%%%%%%%%%%%%%
\section{Metric coefficient in the presence of DM plasma}
In this appendix, we remind ourselves the functional dependence of the metric coefficients for the case of a spherically symmetric blackhole surrounded by an isotropic perfect fluid. The following treatment will be applicable to the DM plasma case we study, where the total mass of the DM is negligible compared to the black hole at the centre, and the pressure is assumed to be vanishing to leading order.

Though the general treatment of such scenarios is well known (see for example\,\cite{weinberg1972gravitation}), we wish to elucidate a few points about how certain metric coefficients are affected by the presence of the DM plasma, and clarify certain ambiguities in the literature, concerning under what assumptions certain identities are applicable. As we mentioned in the text, note in particular that the angular size of the black hole shadow only depends on the $g_{tt}$ component of the metric, while the null ray trajectories depend in addition on $g_{rr}$. Thus, it is for instance important to understand under what limits the two coefficients may be approximated to be inverses of each other. We take $G_{\text{\tiny{N}}}=c=1$ in this appendix.

The most general line element for a static, spherically symmetric spacetime may be taken as
\begin{equation}\label{eq:A2le}
ds^2=-f(r)dt^2+\frac{dr^2}{g(r)}+r^2d\Omega^2\; .
\end{equation}
Let $f(r)=e^{a(r)}$ and $g(r)=e^{-b(r)}$, without loss of generality. The energy-momentum tensor for an isotropic perfect fluid is
\begin{equation}\label{eq:A2emt}
T_{\mu\nu}=(\rho(r)+p(r))U_\mu U_\nu +p(r)g_{\mu\nu}\;. 
\end{equation}
where $\rho(r)$ and $p(r)$ are the proper energy density and the proper pressure, assumed to be functions of just the radial distance $r$. As we are interested in static solutions, we can define a time-like four velocity. The normalised (i.e. satisfying $U^\mu U_\mu=-1$) time-like four velocity is
\begin{equation}\label{eq:A2fv}
U^\mu=(e^{-\frac{a(r)}{2}},0,0,0)\;.
\end{equation}

Using Eq.\,(\ref{eq:A2fv}) in Eq.\,(\ref{eq:A2emt}), gives
\begin{equation}\label{eq:A2emt1}
T^\mu_\nu=Diag(-\rho,p,p,p)\;.
\end{equation}
On then substituting Eq.\,(\ref{eq:A2emt1}) into the Einstein field equation 
\begin{equation}\label{eq:A2ee}
G^\mu_\nu=R^\mu_\nu-\frac{1}{2}R\delta^\mu_\nu = 8\pi T^\mu_\nu \;,
\end{equation}
we get the equations for the relevant components
\begin{equation}
\label{eq:A2ee1tt}
G^t_t=-\frac{e^{-b(r)}}{r^2}\left(rb'(r)+e^{b(r)}-1\right)=-8\pi\rho \;,
\end{equation}
\begin{equation}
\label{eq:A2err}
G^r_r=\frac{e^{-b(r)}}{r^2}\left(ra'(r)-e^{b(r)}+1\right)=8\pi p \;,
\end{equation}
\begin{equation}
\label{eq:A2eettpp}
G^\theta_\theta=\frac{e^{-b(r)}}{4r}\left(2a'(r)+r(a'(r))^2-2b'(r)-ra'(r)b'(r)+2ra''(r)\right)=8\pi p\;.
\end{equation}

Following \cite{weinberg1972gravitation}, we solve Eq.\,(\ref{eq:A2ee1tt}) and  obtain 
\begin{equation}\label{eq:A2bs}
e^{b(r)}=\left(1-\frac{2}{r}m(r)\right)^{-1}\;,
\end{equation}
where
\begin{equation}\label{eq:A2m}
m(r)=\int_0^r\,dr'~4\pi\rho(r') r'^2\;,
\end{equation}
is the total mass enclosed in a region of radius $r$. 

Eqs.\,(\ref{eq:A2bs}) and (\ref{eq:A2m}) will give us the $g_{rr}$ component of the metric. For computing the $g_{tt}$ component we use Eqs.\,(\ref{eq:A2err}) and (\ref{eq:A2m}) to obtain
\begin{equation}\label{eq:A2as}
a'(r)=\frac{2(m(r)+4\pi p r^3)}{r^2\left(1-\frac{2}{r}m(r)\right)}\;.
\end{equation}

The DM plasma component, as well as the total DM, will be assumed to be pressure-less ($p_{\text{\tiny{DM}}}=0$) and collision-less in our study. As already mentioned, in the very special case\,\cite{Kiselev:2003ah,Li:2012zx} with a quintessence-like DM equation of state $P_r\simeq -\rho$, analytic solutions for the Einstein field equations are known, satisfying $g_{tt}=g_{rr}^{-1}$. Our focus is on the physically viable mCP parameter space, and the DM plasma components that arise from these sectors. In the very low-mass mCP regime the DM plasma pressure  may be important, and effect the metric coefficients. Nevertheless, the deviations on the photon sphere and shadow radii due to such corrections, from the relevant DM plasma fluid polytropic equation of state, will only cause an even larger deviation than the pressure-less case we consider. Therefore, as commented in the text, our study still gives a conservative estimate for the DM plasma induced deviations.

With this assumption, Eq.\,(\ref{eq:A2as}) then gives 
\begin{equation}\label{eq:A2asp}
a'(r)=\frac{2m(r)}{r^2\left(1-\frac{2}{r}m(r)\right)}\;.
\end{equation}

When we consider the case of a solitary Schwarzschild blackhole, we would have
\begin{equation}\label{eq:A2sd}
\rho(r)=\rho_{\text{\tiny  Sch.}}(r)\equiv\frac{M_{\text{\tiny{BH}}}}{4\pi r^2}\delta^3(r)\;,
\end{equation}
and the corresponding mass function from Eq.\,(\ref{eq:A2m}) would be
\begin{equation}
M_{\text{\tiny  Sch.}}(r)=M_{\text{\tiny{BH}}}\,,\hquad\text{for } r>0\;.
\end{equation}
Then, using Eqs.\,(\ref{eq:A2bs}) and (\ref{eq:A2asp}), and taking the exterior spacetime to be asymptotically flat, we would obtain as expected
\begin{equation}\label{eq:A2sm}
\left. e^{a_(r)}\right|_{\text{\tiny  Sch.}}=\left.e^{-b(r)}\right|_{\text{\tiny  Sch.}}=\left(1-\frac{2}{r}M_{\text{\tiny{BH}}}\right)\;.
\end{equation} 

Let us now consider the case of a spherically symmetric DM distribution, that has the DM plasma as a component, around a Schwarzschild black hole. As we already mentioned, we will assume that the total mass of the DM, over the finite DM plasma region under consideration, is negligible compared to the central black hole mass. The density function over this finite range would be given by
\begin{equation}\label{eq:A2rhod}
\rho(r)=\frac{M_{\text{\tiny{BH}}}}{4\pi r^2}\delta^3(r) + \Theta(r-R_{{\text{\tiny  Sch.}}})\rho_{\text{\tiny{DM}}}\;.
\end{equation}
Here, $\Theta(x)$ is the Heaviside function. Using Eq.\,(\ref{eq:A2m}), we then obtain the mass function for this Schwarzschild-DM system (Sch.-DM) as 
\begin{equation}
M_{{\text{\tiny Sch.-DM}}}(r)\equiv M_{\text{\tiny{BH}}} +M_{\text{\tiny{DM}}}(r) \;,
\end{equation}
where $M_{\text{\tiny{DM}}}(r)=\int_{R_{{\text{\tiny  Sch.}}}}^rdr~4\pi r^2 \rho_{\text{\tiny{DM}}}$ is the mass of the DM enclosed in a sphere of radius r. 

Using Eq.\,(\ref{eq:A2bs}), the coefficient $g_{rr}$ in the metric will be
\begin{equation}
\left.e^{b(r)}\right|_{{\text{\tiny Sch.-DM}}}=\left(1-\frac{2}{r}(M_{\text{\tiny{BH}}}+M_{\text{\tiny{DM}}}(r))\right)^{-1}\;.
\end{equation}
For the $g_{tt}$ we would have
\begin{equation}
\label{eq:gttfull}
\left.e^{a(r)}\right|_{{\text{\tiny Sch.-DM}}}=\exp \left[ \int_0^r \,dr'~ \frac{2\,\left( M_{\text{\tiny{BH}}} +M_{\text{\tiny{DM}}}(r') \right)}{r'^2\left[1-\frac{2}{r'}\left( M_{\text{\tiny{BH}}} +M_{\text{\tiny{DM}}}(r') \right)\right]}\right]
\end{equation}
This is the general expression for the metric coefficients. For the $g_{tt}$ component, in the limit of the total DM mass being much smaller than the mass of the blackhole, we may expand Eq.\,(\ref{eq:gttfull}) perturbatively in $M_{\text{\tiny{DM}}}(r)/M_{\text{\tiny{BH}}}$. To leading order, using Eq.\,(\ref{eq:gttfull}), this would give
\begin{equation}
\left.e^{a(r)}\right|_{{\text{\tiny Sch.-DM}}}\simeq\left(1-\frac{2}{r}(M_{\text{\tiny{BH}}}+M_{\text{\tiny{DM}}}(r))\right)\;,
\end{equation}
with higher order correction terms involving integrals over powers of $M_{\text{\tiny{DM}}}(r)/M_{\text{\tiny{BH}}}$. To leading order we therefore have,
\begin{equation}\label{eq:A2grrgtt}
\left.e^{a(r)}\right|_{{\text{\tiny Sch.-DM}}}\simeq \left. e^{-b(r)}\right|_{{\text{\tiny Sch.-DM}}}=\left(1-\frac{2}{r}(M_{\text{\tiny{BH}}}+M_{\text{\tiny{DM}}}(r))\right)\; .
\end{equation}

%%%%%%%%%%%%%%%%%%%%%%%%%%%%%%%%%%%%%%%%%%%%
\section{Propagation of light through DM plasma in curved spacetimes}
In this appendix, we will briefly summarise the incorporation of  DM plasma effects, when the DM plasma is treated as a two-component fluid. Specifically, we would like to study the influence of DM plasma on the propagation of light in curved spacetimes. The plasma is assumed to be composed of two charged species $\tilde{e}$ and $\tilde{p}$ with charges -$\epsilon q_e$ and $\epsilon q_e$. They have masses $m_{\tilde{e}}$ and $m_{\tilde{p}}$ respectively. 

We will consider the DM plasma to be a cold, non-magnetized, pressure-less fluid,  and will follow the well-known formalism of a two-component fluid\,\cite{Synge:1960ueh, breuer1980propagation,breuer1981propagation,perlick2000ray}. We are interested in the propagation of electromagnetic waves in the DM plasma, and the presence of pressure in the DM plasma does not in any case affect the dispersion relation for transverse waves. They only have an effect on the longitudinal waves in the plasma, and effect their dispersion relations (see for instance,\,\cite{1998pfp..book.....C,thorne2017modern}). Motivated by conventional baryonic plasmas, the positive ions are sometimes assumed to be very heavy, in comparison to the electrons\,\cite{perlick2000ray}. However, in the context of the DM plasma we will consider the general case where ${\tilde{p}}$ is not necessarily much heavier than ${\tilde{e}}$. We then find that the dynamics of this system is governed by a system of non-linear first order differential equations, for $F_{ab}$, $n_{\left(\tilde{e},\tilde{p}\right)}$ and $U^{a}_{\left(\tilde{e},\tilde{p}\right)}$, of the form
\begin{eqnarray}
	\label{eq:A1me1}
	\partial_{\left[a\right.}F_{\left.bc\right]}&=&0 \;,\\
	\label{eq:A1me2}
	\nabla_b F^{ab}&=&\epsilon q_e n_{\tilde{p}} U^a_{\tilde{p}}-\epsilon q_e n_{\tilde{e}} U^a_{\tilde{e}} \;, \\
	\label{eq:A1lfe}
	m_{\tilde{e}}U^{b}_{\tilde{e}}\nabla_{b}U_{\tilde{e}}^{a}&=&-\epsilon q_e F^{a}_{b}U_{\tilde{e}}^{b} \;, \\
	\label{eq:A1lfp}
	m_{\tilde{p}}U_{\tilde{p}}^{b}\nabla_{b}U_{\tilde{p}}^{a}&=&\epsilon q_e F^{a}_{b}U_{\tilde{p}}^{b} \;, \\
	\label{eq:A1ce}
	\nabla_a\left(n_{\left(\tilde{e},\tilde{p}\right)}U^{a}_{\left(\tilde{e},\tilde{p}\right)}\right)&=&0 \;, \\
	\label{eq:A1ttm}
	g_{ab}U^{a}_{\left(\tilde{e},\tilde{p}\right)}U^{b}_{\left(\tilde{e},\tilde{p}\right)}&=&-1 \;.
\end{eqnarray}
Here, Eqs.\,(\ref{eq:A1me1}) and (\ref{eq:A1me2}) are the Maxwell equations for the electromagnetic field strength tensor $F_{ab}$, with $U^{a}_{\left(\tilde{e},\tilde{p}\right)}$ and $n_{\left(\tilde{e},\tilde{p}\right)}$ as the four-velocity and number density for the respective components. $\nabla_{a}$ are the covariant derivatives. Eqs.\,(\ref{eq:A1lfe}) and (\ref{eq:A1lfp}) are the equations of motion, or the force equations, for the respective fluid components. These are combined forms of the Euler equation and the Lorentz force law. Eq.\,(\ref{eq:A1ce}) is just the equation of continuity. The metric $g_{ab}$ in Eq.\,(\ref{eq:A1ttm}) specifies the structure of the spacetime.

 It has been shown\,\cite{breuer1980propagation} that the system of equations Eqs.\,(\ref{eq:A1me1})-(\ref{eq:A1ttm}) are stable with respect to linearisation. Therefore, one can solve these equations by considering a perturbation about a background solution. Therefore, let us consider\footnote[1]{For charge neutrality, $n^{(0)}\equiv n_{\tilde{e}}^{(0)}=n_{\tilde{p}}^{(0)}$ }
 \begin{eqnarray}
 	F_{ab}(x)&=&0+{F}_{ab}^{(1)}(x)\;,\\
 	n_{\left(\tilde{e},\tilde{p}\right)}(x)&=&n^{(0)}(x)+{n}_{\left(\tilde{e},\tilde{p}\right)}^{(1)}(x)\;,\\ U^{a}_{\left(\tilde{e},\tilde{p}\right)}(x)&=&U^{(0)a}_{\left(\tilde{e},\tilde{p}\right)}(x)+{U}^{(1)a}_{\left(\tilde{e},\tilde{p}\right)}(x)\;.
 \end{eqnarray}
 The background solution will satisfy the equations
     \begin{eqnarray}
     	\label{eq:A1bsme2}
      U^{(0)a}_{\tilde{p}}-U^{(0)a}_{\tilde{e}}&=&0 \;, \\
     	\label{eq:A1bslf}
      U^{(0)b}_{\left(\tilde{e},\tilde{p}\right)}\nabla_{b}U_{\left(\tilde{e},\tilde{p}\right)}^{(0)a}&=&0 \;, \\
     	\nabla_a\left(n^{(0)}U^{(0)a}_{\left(\tilde{e},\tilde{p}\right)}\right)&=&0 \;, \\
     	\label{eq:A1bsttm}
     	g_{ab}U^{(0)a}_{\left(\tilde{e},\tilde{p}\right)}U^{(0)b}_{\left(\tilde{e},\tilde{p}\right)}&=&-1 \;.
     \end{eqnarray}
 We define $U^{(0)a}\equiv U^{(0)a}_{\tilde{p}}=U^{(0)a}_{\tilde{e}}$ (from Eq.\,(\ref{eq:A1bsme2})). We deduce that the perturbed solution satisfies the dynamical equations
 \begin{eqnarray}
 	\label{eq:A1psme1}
 	\partial_{\left[a\right.}F^{(1)}_{\left.bc\right]}&=&0 \;,\\
 	\label{eq:A1psme2}
 	\nabla_b F^{(1)ab}&=&\epsilon q_e n^{(0)} \left[\left(U^{(1)a}_{\tilde{p}}-U^{(1)a}_{\tilde{e}}\right) +U^{(0)a} \left(n^{(1)}_{\tilde{p}}-n^{(1)}_{\tilde{e}}\right)\right] \;, ~~~~~~~~\\
 	\label{eq:A1pslfe}
 	m_{\tilde{e}}U^{(0)b}\nabla_{b}U_{\tilde{e}}^{(1)a}+m_{\tilde{e}}U^{(1)b}_{\tilde{e}}\nabla_{b}U^{(0)a}&=&-\epsilon q_e F^{(1)a}_{b}U^{(0)b} \;, \\
 	\label{eq:A1pslfp}
 	m_{\tilde{p}}U^{(0)b}\nabla_{b}U_{\tilde{p}}^{(1)a}+m_{\tilde{p}}U^{(1)b}_{\tilde{p}}\nabla_{b}U^{(0)a}&=&\epsilon q_e F^{(1)a}_{b}U^{(0)b} \;, \\
 	\label{eq:A1psce}
 	\nabla_a\left(n^{(0)}U^{(1)a}_{\left(\tilde{e},\tilde{p}\right)}+n^{(1)}_{\left(\tilde{e},\tilde{p}\right)}U^{(0)a}\right)&=&0 \;, \\
 	\label{eq:A1psttm}
 	g_{ab}U^{(0)a}U^{(1)b}_{\left(\tilde{e},\tilde{p}\right)}&=&0 \;.
 \end{eqnarray}
 As we are interested in the dynamics of the electromagentic field, we intend to find the equation for $F^{(1){ab}}$. Using Eqs.\,(\ref{eq:A1ttm}), (\ref{eq:A1psme2}), and (\ref{eq:A1psttm}), we obtain a relation for $n^{(1)}_{\left(\tilde{e},\tilde{p}\right)}$  and $U^{(1)a}_{\left(\tilde{e},\tilde{p}\right)}$ in terms of $F^{(1){ab}}$ of the form,
 \begin{eqnarray}
 	\label{eq:A1sn}
 	\epsilon q_e \left(n^{(1)}_{\tilde{p}}-n^{(1)}_{\tilde{e}}\right)&=&	-U^{(0)a}\nabla_b F^{(1)ab} \, ,\\
 	\label{eq:A1sU}
 	\epsilon q_e n^{(0)}\left(U^{(1)a}_{\tilde{p}}-U^{(1)a}_{\tilde{e}}\right)&=&	\nabla_b F^{(1)cb}\left(\delta^a_c+U^{(0)a}U_{(0)c}\right) \, .
 \end{eqnarray}
 The dynamical equation for $F^{(1)_{ab}}$ can be obtained by substituting Eq.\,(\ref{eq:A1sU}) in the difference of Eq.\,(\ref{eq:A1pslfp}) and Eq.\,(\ref{eq:A1pslfe}). This gives
 
 	\begin{eqnarray}
 		U^{(0)b}\left(\delta^a_c+U^{(0)a}U_{(0)c}\right)\nabla_b\nabla_d F^{(1)cd} &+&\left(\nabla_{b}U^{(0)b}\left(\delta^a_c+U^{(0)a}U_{(0)c}\right)+\nabla_{c}U^{(0)a}\right)\nabla_d F^{(1)cd}\nonumber\\
 		\label{eq:A1sF}
 		&=&\tilde{\omega}^2_p U^{(0)a} F^{(1)a}_b\,.
 	\end{eqnarray}
 Here, $\tilde{\omega}^2_p=\epsilon^2q_e^2 n^{(0)}/\mu$ is the plasma frequency\footnote[2]{In S.I. units, $\tilde{\omega}^2_p=\epsilon^2 q_e^2 n^{(0)}/(\epsilon_o\mu)$  } due to $\tilde{e}$ and $\tilde{p}$. $\mu=m_{\tilde{e}}m_{\tilde{p}}/(m_{\tilde{e}}+m_{\tilde{p}})$ is the reduced mass. 
 
 One may express $F^{(1){ab}}$ in terms of the perturbed potential $A^{(1)}_a$  as $F^{(1){ab}}=\nabla_{_{\left[a\right.}}A_{\left.b\right]}^{(1)}$. With this substitution, Eq.\,(\ref{eq:A1sF}) takes the form
 \begin{eqnarray}
 \mathcal{D}^{af}A^{(1)}_f\equiv~U^{(0)b}\left(\delta^a_c+U^{(0)a}U_{(0)c}\right)\nabla_b\left(\nabla^f\nabla^c-g^{fc}\nabla^d\nabla_d\right)A^{(1)}_f &+&\nonumber\\
 \label{eq:A1sA}
 \left(\nabla_{b}U^{(0)b}\left(\delta^a_c+U^{(0)a}U_{(0)c}\right)+\nabla_{c}U^{(0)a}\right)\nabla_d \left(\nabla^f\nabla^c-g^{fc}\nabla^d\nabla_d\right)A^{(1)}_f&&\\
 -\tilde{\omega}^2_p \left(U^{(0)f}\nabla^a-g^{af}U^{(0)b}\nabla_b\right) A^{(1)}_f&=&0\,.\nonumber
 \end{eqnarray}
 Furthermore, it is assumed that the potential $A^{(1)}_a$ satisfies the Landau gauge condition
 \begin{equation}
 \label{eq:A1lg}
 	A^{(1)}_aU^{(0)a}=0 \;.
 \end{equation}
 Eq.\,(\ref{eq:A1sA}) along with Eq.\,(\ref{eq:A1lg}) determines the solution for the system. 
  In \cite{breuer1980propagation,breuer1981propagation}, they have considered a two parameter family of approximate plane wave ansatz for $A^{(1)}_a$ i.e.
  \begin{equation}
  \label{eq:A1aa}
  A^{(1)}_a (x,\upsilon,\xi)=\frac{\upsilon}{\xi}\text{Re}\left(e^{iS(x_o+\xi(x-x_{o}))/\alpha}~a^{(1)}_a (x_o+\xi(x-x_{o}))\right)\;,
  \end{equation}
  and obtained the eikonal equations  for the system as
  \begin{equation}
  	U^{(0)a}\partial_a S\left(-(U^{(0)b}\partial_b S)^2+\tilde{\omega}^2_p\right)\left(\partial_c S\partial^c S+\tilde{\omega}^2_p\right)=0\;.
  \end{equation}

 As mentioned in~\cite{breuer1980propagation,breuer1981propagation}, the different solutions for these eikonal equations can be obtained from that of one of the three partial Hamiltonians below
 \begin{eqnarray}
 	\label{eq:A1pe1}
 	 H_{1}(x,p)&\equiv&U^{(0)a}p_{a}=0\,,\\
 	\label{eq:A1pe2}
 	H_{2}(x,p)&\equiv& \frac{1}{2}\left(-U^{(0)a}U^{(0)b}p_{a}p_{b}+\tilde{\omega}^2_{p}\right)=0\,,\\
 	\label{eq:A1pe3}
 	H_{3}(x,p)&\equiv& \frac{1}{2}\left(g^{ab}p_{a}p_{b}+\tilde{\omega}^2_{p}\right)=0\,.
 \end{eqnarray}
 
 The solution of the Hamiltonian $H_{1}(x,p)=0$ is irrelevant because it will result in a vanishing frequency for the electromagnetic wave. The solution of the Hamiltonian $H_{2}(x,p)=0$  describes the longitudinal mode of plasma oscillation. In the study, as we are interested in the passage of electromagnetic wave through the DM plasma. Therefore, the Hamiltonian $H_3(x,p)$, describing the transverse mode, will be the relevant expression of interest to us. This is the Hamiltonian that we will therfore take to analyse the light ray trajectories through the DM plasma. From $H_3(x,p)$, as $g^{ab}p_{a}p_{b}<0$, we see that the photon rays will follow a timelike trajectory through the DM plasma.

%%%%%%%%%%%%%%%%%%%%%%%%%%%%%%%%%%%%%%%%%%%%%%%%%%%%%%%%%%%%%
\bibliographystyle{JHEP.bst}
\bibliography{mCPBHShadow} 
%%%%%%%%%%%%%%%%%%%%%%%%%%%%%%%%%%%%%%%%%%%%%%%%%%%%%%%%%%%%%

\end{document}